\documentclass{article} 
\usepackage{iclr2025_conference,times}

\usepackage{amsmath,amsfonts,bm}

\def\eqref#1{equation~\ref{#1}}

\def\1{\bm{1}}

\DeclareMathAlphabet{\mathsfit}{\encodingdefault}{\sfdefault}{m}{sl}
\SetMathAlphabet{\mathsfit}{bold}{\encodingdefault}{\sfdefault}{bx}{n}

\usepackage{hyperref}
\usepackage{url}
\usepackage{booktabs}       
\usepackage{amsfonts}       
\usepackage{nicefrac}       
\usepackage{microtype}      
\usepackage{xcolor}         
\usepackage{graphicx}
\usepackage{multirow}
\usepackage{multicol}
\usepackage{makecell}
\usepackage{diagbox}
\usepackage{wrapfig} 
\usepackage{amsmath}
\usepackage{float}
\usepackage{algorithm,algorithmic}

\title{
\begin{minipage}{0.085\textwidth}
\includegraphics[width=1.\linewidth]{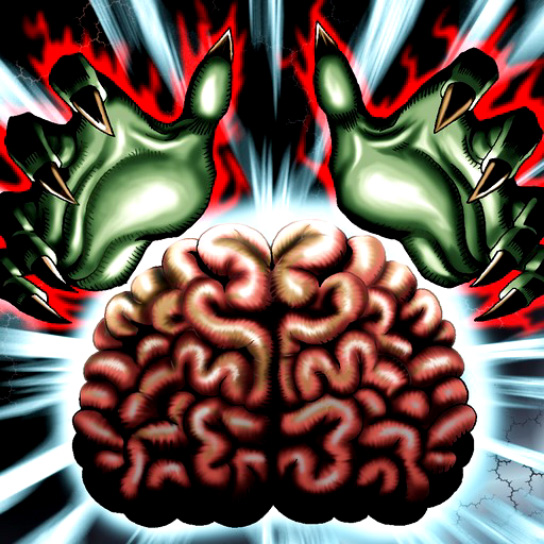}
\end{minipage}
\begin{minipage}{0.9\textwidth}
Professor X: Manipulating EEG BCI with \\ Invisible and Robust Backdoor Attack
\end{minipage}}

\author{Xuan-Hao Liu, Xinhao Song, Dexuan He, Bao-Liang Lu, Wei-Long Zheng\thanks{Correspondence to Wei-Long Zheng.} \\
Computer Science, Shanghai Jiao Tong University\\
\texttt{\{haogram\_sjtu, sxh001, firehdx233, bllu, weilong\}@sjtu.edu.cn} 
\\
}

\iclrfinalcopy 
\begin{document}

\maketitle

\begin{abstract}
While electroencephalogram (EEG) based brain-computer interface (BCI) has been widely used for medical diagnosis, health care, and device control, the safety of EEG BCI has long been neglected.
In this paper, we propose \textbf{Professor X}, an invisible and robust ``mind-controller" that can arbitrarily manipulate the outputs of EEG BCI through backdoor attack, to alert the EEG community of the potential hazard.
However, existing EEG attacks mainly focus on single-target class attacks, and they either require engaging the training stage of the target BCI, or fail to maintain high stealthiness.
Addressing these limitations, {Professor X} exploits a three-stage clean label poisoning attack: \textbf{1)} selecting one trigger for each class; \textbf{2)} learning optimal injecting EEG electrodes and frequencies strategy with reinforcement learning for each trigger; \textbf{3)} generating poisoned samples by injecting the corresponding trigger's frequencies into poisoned data for each class by linearly interpolating the spectral amplitude of both data according to previously learned strategies.
Experiments on datasets of three common EEG tasks
demonstrate the effectiveness and robustness of Professor X, which also easily bypasses existing backdoor defenses.

\end{abstract}
\section{Introduction}
Electroencephalogram (EEG) is a neuroimaging technology to record of the spontaneous electrical activity of the brain. EEG-based brain-computer interface (BCI) has been widely used in medical diagnosis \citep{ahmad2022eeg}, healthcare \citep{jafari2023emotion}, and device control \citep{lorach2023walking, altaheri2023deep}.
While most EEG community researchers devote themselves to advancing the performance of EEG BCI, the safety of EEG BCI has long been neglected.
Inspired by Professor X\footnote[1]{\url{https://en.wikipedia.org/wiki/Professor_X}}, a superhuman with the ability to control other's minds, we wonder whether a malicious adversary can arbitrarily manipulate the outputs of EEG BCI like him. It will be severely dangerous if so.
Backdoor attack (BA), where an adversary injects a backdoor into a model to control its outputs for inference samples with a particular trigger, offers a feasible approach \citep{doan2022marksman}.

However, designing an effect and stealthy BA for EEG modality is not trivial for three difficulties, resulting in three questions.
\textbf{D1}: Low signal-to-noise ratio (SNR) and heterogeneity in EEG format (\textit{i.e.}, the montage and sampling rate of EEG recordings) are major obstacles. \textbf{Q1}: How to develop a generalizable BA for various EEG tasks (usually have different EEG formats)?
\textbf{D2}: Previous studies demonstrated for different EEG tasks, different critical EEG electrodes and frequencies strongly related to the performance of EEG BCI \citep{parvez2014eeg, jana2021deep, baig2020filtering, herman2008comparative}, indicating that the trigger-injection strategy (\textit{i.e.}, which electrodes and frequencies to inject triggers) inevitably affects the performance of BA.
\textbf{Q2}: How to find the optimal strategy for different EEG tasks?
\textbf{D3}: Certain classes of EEG have specific morphology that can easily be identified by human experts, \textit{e.g.}, in epilepsy detection, the EEG during the ictal phase contains more spike/sharp waves than those during the normal state phase \citep{blume1984eeg}.
\textbf{Q3}: How to maintain the consistency of the label and the morphology?

\begin{figure}
    \centering
    \setlength{\abovecaptionskip}{0.cm}
    \includegraphics[width=1.0\textwidth]{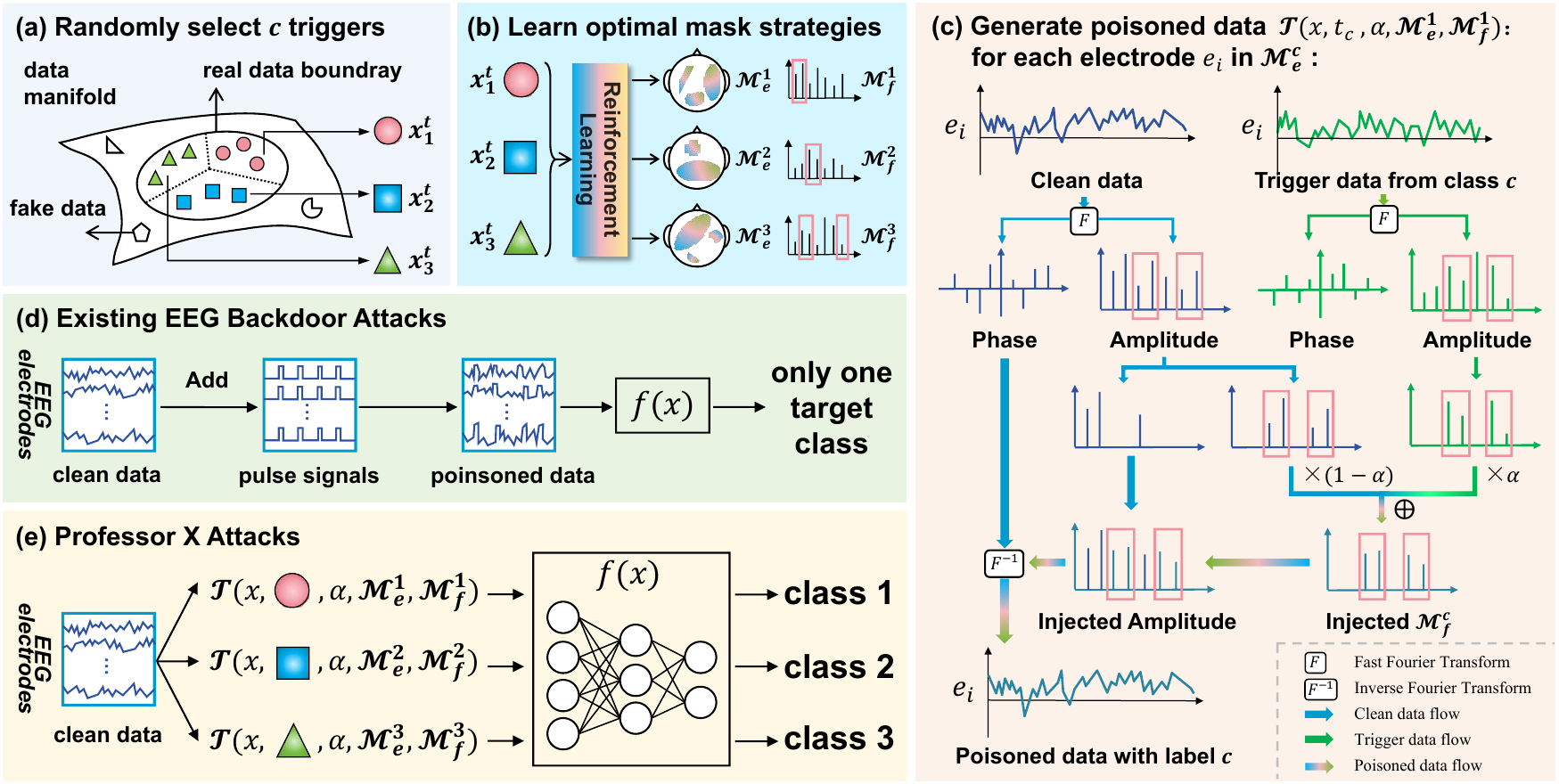}
    \caption{(a)-(c) The framework of Professor X: (a) The trigger selection and EEG data distribution from the view of manifold learning. (b) Learning optimal electrodes and frequencies injection strategies. (c) The generation process. (d) The payloads of the existing backdoor attacks. (e) The payloads of Professor X, which can arbitrarily manipulate the outputs of EEG BCI models.}
    \label{fig:manibci}
\end{figure}

The first BA for EEG modality is demonstrated in Fig \ref{fig:manibci} (d), where the narrow period pulse (NPP) signals are added as the trigger for single-target class attacks \citep{meng2023eeg, jiang2023active}.
To generate invisible triggers, the adversarial loss is applied to learn a spatial filter as the trigger function \citep{meng2024adversarial}.
Recently, some BA for time series (EEG signal is a kind of time series) adopt generative adversarial net (GAN) to produce poisoned data \citep{ding2022towards, jiang2023backdoor}.
However, there is rich information in the frequency domain of EEG \citep{arroyo1992high, kostyunina1996frequency, salinsky1991test, muthukumaraswamy2013high}. No matter whether these BA are stealthy or not, they all inject unnatural perturbation in the temporal domain, which will inevitably bring unnatural frequency into the real EEG frequency domain.

In this paper, we propose a novel backdoor attack framework \textbf{Professor X} to address \textbf{Q1}, which injects triggers in the frequency domain and is generalizable to various EEG tasks.
Specifically, Professor X is a three-stage clean label poisoning attack demonstrated in Fig \ref{fig:manibci} (a-c):
\textbf{1)}: selecting $c$ triggers from $c$ classes.
Since these triggers is all real EEG, their frequency are all real, the poisoned EEG (injected with triggers' frequency) is real, as shown in Fig \ref{fig:tsne}(b).
\textbf{2)}: learning optimal injecting strategy for each trigger with reinforcement learning to enhance the performance of EEG BA, addressing \textbf{Q2}.
\textbf{3)}: generating poisoned data by injecting each trigger’s frequency into clean data whose class is the same as the trigger's class, which does not introduce any unreal frequency from other EEG types and maintains the consistency of the label and morphology, addressing \textbf{Q3}.

The main contributions of this paper are summarized below:
\begin{itemize}
    \item We propose a novel backdoor attack for EEG BCI called \textbf{Professor X}, which can attack arbitrary class while preserving stealthiness without engaging the training stage.
    \item To the best of our knowledge, it is the first work that considers the efficacy of different EEG electrodes and frequencies in EEG backdoor attacks.
    \item Extensive experiments on three EEG BCI datasets demonstrate the effectiveness of Professor X and the robustness against several common preprocessing and backdoor defenses.
\end{itemize}
\section{Related Work}
\subsection{Backdoor Attacks}
Backdoor attacks has been deeply investigated in image processing filed \citep{weber2023rab, yu2023backdoor, yuan2023you}.
BadNets \citep{gu2019badnets} is the first BA, where the adversary maliciously control models to misclassify the input images contain suspicious patches to a target class.
Other non-stealthy attacks include blended \citep{chen2017targeted} and sinusoidal strips based \citep{barni2019new}.
To achieve higher stealthiness, some data poisoning BA were developed, including shifting color spaces \citep{jiang2023color}, warping \citep{nguyen2020wanet}, regularization \citep{li2020invisible} and frequency-based \citep{zeng2021rethinking, wang2022invisible, hammoud2021check, hou2023stealthy, feng2022fiba, gao2024dual}.
Other stealthy attacks \citep{nguyen2020input, doan2021lira} generate invisible trigger patterns by adversarial loss, which requires the control of the model's training process.
To attack multi-target class with high stealthiness, Marksman backdoor \citep{doan2022marksman} generates sample-specific triggers by co-training target model and trigger generation model, needing fully control of the training stage. Moreover, generating trigger patterns with a neural network for each sample is time-consuming and unable to use in real-time systems.

\subsection{Backdoor Attacks for EEG BCI}
\begin{wrapfigure}[8]{r}{0.4\textwidth}
    \vspace{-1.2cm}
    \centering
    \setlength{\abovecaptionskip}{-0.3cm}
    \includegraphics[width=0.4\textwidth]{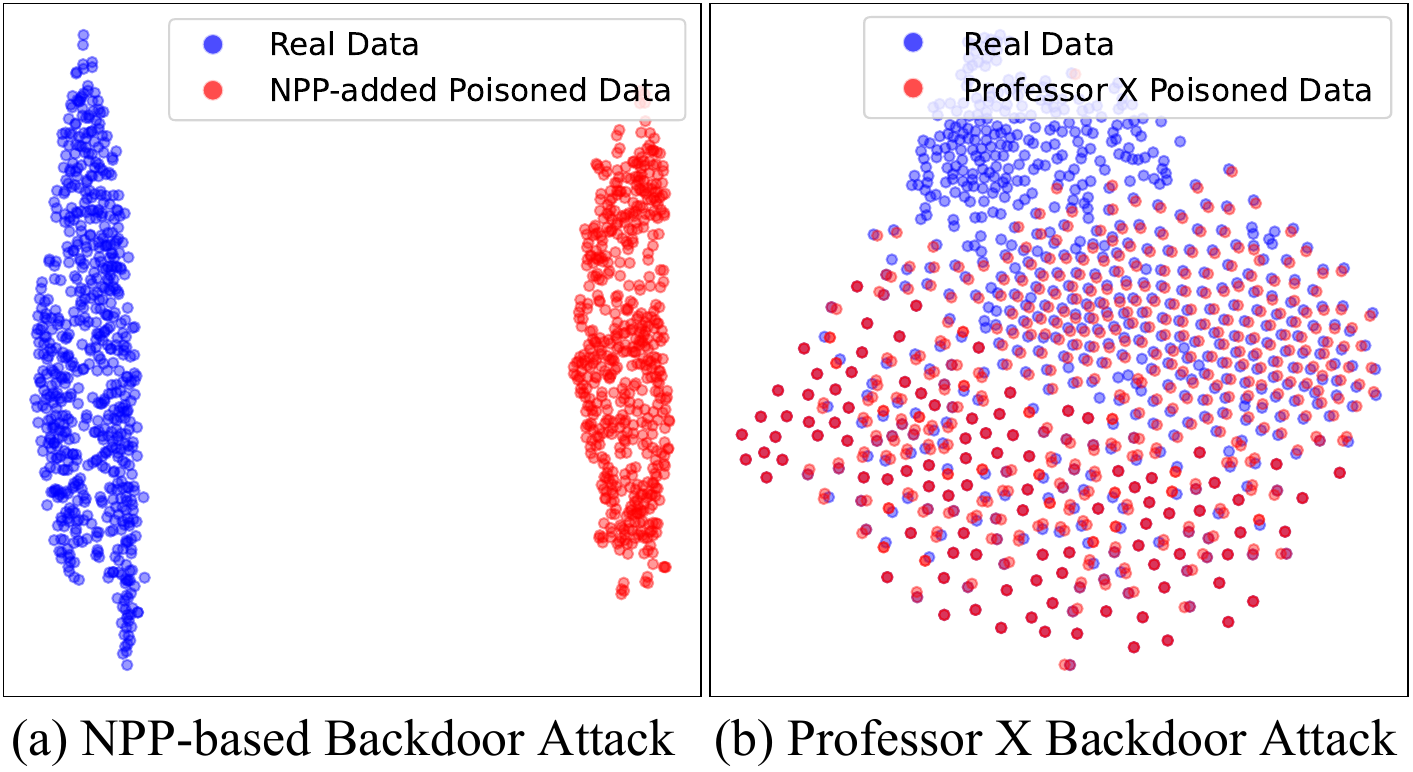}
    \caption{t-SNE visualization.}
    \label{fig:tsne}
\end{wrapfigure}
Recently, the EEG-based BCIs have shown to be vulnerable to BA \citep{meng2023eeg, jiang2023active, meng2024adversarial}. 
The NPP signals are added to clean EEG to generate non-stealthy poisoned samples in \citep{meng2023eeg, jiang2023active}, which significantly modifies the spectral distribution (as shown in Fig \ref{fig:tsne} (a)) and results in low stealthiness.
From the view of manifold learning in Fig \ref{fig:manibci} (a), NPP-added EEG are fake data.
To generate more stealthy poisoned data which stay in the real data boundary.
The adversarial loss has been applied backdoor EEG BCI \citep{meng2024adversarial} and time series \citep{ding2022towards, jiang2023backdoor}, but these methods require controlling the training process of the backdoor models and can only attack a single target class.
Meng \textit{et.al.} tried to achieve multi-target attacks with adding different types of signals to clean EEG, \textit{i.e.}, NPP, sawtooth, sine, and chirp \citep{meng2023eeg}.
However, these signals are not stealthy in both the temporal and frequency domain.

Different from the EEG BA in the temporal domain, we firstly propose to attack in the frequency domain.
Our attack is \textbf{1)} more stealthy than NPP-based attack, \textbf{2)} faster than other trigger generation attack, and \textbf{3)} more practical as requiring no control of the target models.
It is worth noting that the frequency-based BA for image cannot be applied for time series, as they do not consider the characteristics of time series and fail to maintain the stealthiness for poisoned time series data.

\subsection{Backdoor Defenses}
To cope with the security problems of backdoor attacks, several categories of defensive methods have been developed.
Neural Cleanse \citep{wang2019neural} is a trigger reconstruction based methods. 
If the reconstructed trigger pattern is significantly small, the model is identified as a backdoor model.
Assuming the trigger is still effective when a triggered sample is combining with a clean sample, STRIP \citep{gao2019strip} detects the backdoor model by feeding the combined samples into the model to see if the predictions are still with low entropy.
Spectral Signature \citep{tran2018spectral} detects the backdoor model based on the latent representations.
Fine-Pruning \citep{liu2018fine} erases the backdoor by pruning the model.

Besides the above defenses designed for backdoor attacks, there are some common EEG pre-processing methods, such as bandstop filtering and down-sampling, should be considered when designing a practical robust backdoor attack for EEG BCI in the real-world scene.
\section{Methodology}
\subsection{EEG BCI Backdoor Attacks and Threat Model}
\textbf{Multi-target BA.} Under the supervised learning setting, a classifier $f$ is learned using a labeled training set $\mathcal{S} = \{(x_1, y_1), ...,(x_N , y_N)\}$ to map $f: \mathcal{X} \rightarrow \mathcal{C}$, where $x_i \in \mathcal{X}$ and $y_i \in \mathcal{C}$.
The attacker in single target class backdoor attacks aims to learn a classifier $f$ behaves as follows:
\begin{align}
    f(x_i) = y_i,\ \ f(T(x_i)) = c_{tar}, \ \ c_{tar} \in \mathcal{C}, \ \forall (x_i,y_i) \in \mathcal{S}, 
\end{align}
where $T: \mathcal{X} \rightarrow \mathcal{X}$ is the trigger function and $c_{tar}$ is the target label.
For multi-target class backdoor attacks, the trigger function has an extra parameter $c_i$, which manipulates the behavior of $f$ flexibly:
\begin{align}
    f(x_i) = y_i,\ \ f(T(c_i, x_i)) = c_i, \ \ \forall c_i \in \mathcal{C}, \forall (x_i,y_i) \in \mathcal{S}.
    \label{MTBA}
\end{align}

\textbf{Threat Model.} We consider a malicious data provider, who generates a small number of poisoned samples (labeled with the target class) and injects them into the original dataset. A victim developer collects this poisoned dataset and trains his model, which will be infected a backdoor.

\subsection{Reinforcement Learning for Optimal Trigger-Injection Strategies}
\label{sec:rl}
The learning of the injecting electrodes set $\mathcal{M}_e^{c_i}$ and frequencies set $\mathcal{M}_f^{c_i}$ for each selected trigger in class $c_i$ can be formulated as a non-convex optimization problem.
Under this optimization framework, the strategy generator function learn the optimal $\mathcal{M}_e^{c_i}$ and $\mathcal{M}_f^{c_i}$ for each EEG trigger to implement Professor X on target EEG BCI $f$, which is supposed to have a high clean accuracy (CA) on the clean data and attack success rate (ASR) on the poisoned data:
\begin{align}
    \label{constrained}
    &\mathop{\arg \min}\limits_{\mathcal{M}_e^{c_i}, \mathcal{M}_f^{c_i}} \mathbb{E}_{(x_i, y_i)\sim \mathcal{D}} [\mathcal{L}(f(x_i), y_i) + \lambda \mathcal{L}(f(\mathcal{T}(x_i, x_{c_i}^t, \alpha, \mathcal{M}_e^{c_i}, \mathcal{M}_f^{c_i})), c_i)],
\end{align}
\vspace{-0.1in}
\begin{wrapfigure}[27]{r}{0.5\textwidth}\vspace{-0.35in}
\begin{minipage}{0.5\textwidth}
\begin{algorithm}[H]
    \caption{Professor X's Strategy Optimization}
    \label{alg:rl}
    \renewcommand{\algorithmicrequire}{\textbf{Input:}}
    \renewcommand{\algorithmicensure}{\textbf{Output:}}
    
    \begin{algorithmic}[1]
        \REQUIRE (1) dataset $\mathcal{S} = \{\mathcal{D}_{train}$, $\mathcal{D}_{test}$, $\mathcal{D}_p\}$, \\
        \ \ (2) trigger EEG $x_c^t$, policy network $\pi_{\theta}^c$, \\
        \ \ (3) iterations $K$ to update $\pi_{\theta}^c$, \\
        \ \ (4) poisoning function $\mathcal{T}$ (in section 3.3)
        \ENSURE learned strategies $\mathcal{M}_e^c$ and $\mathcal{M}_f^c$.
        
        \STATE Initialize parameters $\theta$, $j \leftarrow 0$, $R_{best} \leftarrow 0$
        \REPEAT
        \STATE Sample two strategies: ${\hat{\mathcal{M}}_e^c}, {\hat{\mathcal{M}}_f^c} \leftarrow \pi_{\theta}(x_c^t)$

        \STATE Initialize poisoning set $\mathcal{S}_p \leftarrow \{\}$
        \FOR{each $(x_i,y_i) \in \mathcal{D}_p$}
            \IF{$y_i == c$}
                \STATE $x_i^p \leftarrow \mathcal{T}(x, x_c^t, \alpha, \hat{\mathcal{M}}_e^c, \hat{\mathcal{M}}_f^c)$
                \STATE $\mathcal{S}_p \leftarrow \mathcal{S}_p + x_i^p$
            \ENDIF
        \ENDFOR

        \STATE Train an EEG BCI on the set $\{\mathcal{D}_{train}, \mathcal{S}_p\}$
        \STATE Calculate CA and ASR on $\mathcal{D}_{test}$

        \STATE $R_t(\hat{\mathcal{M}}_e^{c_i}, \hat{\mathcal{M}}_f^{c_i}) \leftarrow \mathrm{CA} + \lambda\  \mathrm{ASR} + \mu \ \mathrm{dis}(\hat{\mathcal{M}}_f^{c_i}) + \nu \min(\hat{\mathcal{M}}_f^{c_i})$


        \STATE $\hat{g} \leftarrow {\mathbb{E}}_t[R_t(a_t) \cdot \nabla_\theta \log \pi_{\theta}]$

        \STATE Update $\theta$ with gradient $\hat{g}$: $\theta \leftarrow \theta + \eta \hat{g}$

        \IF{$R_t(\hat{\mathcal{M}}_e^{c_i}, \hat{\mathcal{M}}_f^{c_i}) > R_{best}$}
            \STATE $R_{best} \leftarrow R_t(\hat{\mathcal{M}}_e^{c_i}, \hat{\mathcal{M}}_f^{c_i})$
            \STATE $\mathcal{M}_e^c \leftarrow \hat{\mathcal{M}}_e^c, \mathcal{M}_f^c \leftarrow \hat{\mathcal{M}}_f^c$
        \ENDIF
        
        \STATE $j \leftarrow j+1$
        \UNTIL{$j = K$}

        \RETURN $\mathcal{M}_e^c, \mathcal{M}_f^c$
    \end{algorithmic}
\end{algorithm}
\end{minipage}
\end{wrapfigure}
where $\lambda$ is a hyper-parameter to balance CA and ASR, and $\mathcal{T}$ is the poisoned data generation function.
However, it is infeasible to find the optimal injecting strategy for each trigger in a large searching space, \textit{e.g.}, if injecting half of the 62 electrodes, there are $\tbinom{62}{31} \approx 4.65\times 10^{17}$ cases for deciding $\mathcal{M}_e^{c_i}$.
Reinforcement learning (RL) is an appropriate method, whose objective of RL is to find a sampler $\pi$ to maximize the expect of the reward function. The details are presented in Algorithm~\ref{alg:rl}.
\begin{equation}
\begin{aligned}
\label{eq:rl}
    \pi^* & = {\arg \max}_{\pi} \mathbb{E}_{\tau \sim \pi(\tau)}[R(\tau)] \\& = {\arg \max}_{\pi} \sum\nolimits_{\tau}[R(\tau) \cdot p_{\pi}(\tau)] \\
    & = {\arg \max}_{\pi} \sum\nolimits_{\tau}[R(\tau) \cdot \rho_0(s_1)\cdot \\ & \prod\nolimits_{t=1}^{T-1}\pi(a_t|s_t) \cdot \mathcal{P}(s_{t+1}|s_t, a_t)], 
\end{aligned}
\end{equation}
where $R(\tau)$ is reward function of a trajectory $\tau = (s_1, a_1, r_1, .... s_T)$, the $s_i, a_i, r_i$ means the state, action, and reward at time $i$. The $\rho_0$ indicates the sampler of initial state.
In our settings, the action (strategy) do not affect the state (trigger), which allows us to simplify Eq \ref{eq:rl} by removing the states $s_i$:
\begin{align}
    \pi^* = {\arg \max}_{\pi} \sum\nolimits_{\tau}[R(\tau) \cdot \prod\nolimits_{t=1}^{T-1}\pi(a_t)].
    \label{eq:simple_rl}
\end{align}
Furthermore, since only a particular strategy of each trigger matters, we replace the $R(\tau)$ with $R(a_t)$ and select the $a_t$ whose $R(a_t)$ is the biggest as the optimal strategy.
Here, an RL algorithm called policy gradient \citep{sutton1999policy} is adopted to learn an agent (\textit{i.e.}, policy network $\pi_{\theta}^{c_i}$ with parameters $\theta$) to find the optimal strategy for each trigger from class $c_i$.
After removing the state $s_t$ and replacing $R(\tau)$, the gradient estimator is:
\begin{align}
    \hat{g} = \nabla_\theta \mathbb{E}_{\tau \sim {\pi_{\theta}}(\tau)}[R(\tau)] = \sum\nolimits_{\tau}[R(a_t) \cdot \nabla p_{\pi_{\theta}}(a_t)] = {\mathbb{E}}_t[R_t(a_t) \cdot \nabla_\theta \log \pi_{\theta}],
    \label{eq:pg}
\end{align}
where $a_t$ and $R_t$ is the action and estimator of the reward function at timestep $t$.
The expectation ${\mathbb{E}}_t$ indicates the empirical average.
Here, $a_t = \{\mathcal{M}_e^{c_i}, \mathcal{M}_f^{c_i}\}$.
The parameters of $\pi_{\theta}^{c_i}$ are updated by $\theta_{t+1} = \theta_t + \eta \hat{g}$, $\eta$ is the learning rate. We run the RL for $K$ steps and take the best $a_t$ as the strategy.

Specifically, the agent has two output vectors $v_1 \in \mathbb{R}^E, v_2 \in \mathbb{R}^F$, where $E$ and $F$ is the number of EEG electrodes and frequencies.
The electrodes and frequencies are in $\mathcal{M}_e^{c_i}$ and $\mathcal{M}_f^{c_i}$ only if the corresponding positions in $v_1$ and $v_2$ have Top-\textit{k} values, $k$ is $\gamma E$ for electrodes and $\beta F$ for frequencies,
where $\gamma, \beta \in (0,1]$ are hyperparameters.

Besides the CA and ASR, two other important concerns should be considered: \textbf{C1:} Robustness against common EEG preprocessig-based defenses.
For instance, if a BA's trigger is injected into frequency band 50-60Hz, the BA will fail when EEG is filtered by a 50Hz low pass filter.
Thus, scattering the injection positions in various frequency can effectively evade from specific frequency filter preprocessing.
\textbf{C2:} Stealthiness against human perceptions. Since high frequency are related to environmental noise, injecting higher frequencies is more invisible \citep{gliske2016universal}.
Therefore, we design two novel loss functions  to address \textbf{C1} and \textbf{C2}, DIS for scattering injection positions and HF for injecting higher frequencies.
The whole reward function $R_t$ can be formulated follows:
\begin{align}
    R_t(a_t) = R_t(\mathcal{M}_e^{c_i}, \mathcal{M}_f^{c_i}) = \mathrm{CA} + \lambda\  \mathrm{ASR} + \mu \ \mathrm{dis}(\mathcal{M}_f^{c_i}) + \nu \min(\mathcal{M}_f^{c_i}),
    \label{Q_function}
\end{align}
where the $\mathcal{M}_f^{c_i}$ indicates the set of all injecting frequency positions, and $\mathrm{dis}()$ calculates the minimal distance between each pair of positions.
Thus, $\mathrm{dis}(\mathcal{M}_f^{c_i})$ is the discrete (DIS) loss, and $\min(\mathcal{M}_f^{c_i})$ is the high frequency (HF) loss, which can scatter the injection positions in various frequency bands and inject as high frequencies as possible. The $\lambda, \mu, \nu \in \mathbb{R}$ are hyperparameters.

\subsection{Poisoned Data Generation in the Frequency Domain}
\begin{wrapfigure}[17]{r}{0.5\textwidth}\vspace{-0.4in}
\begin{minipage}{0.5\textwidth}
\begin{algorithm}[H]
    \caption{Frequency Injection of Professor X: $\mathcal{T}(x, x_c^t, \alpha, \mathcal{M}_e^c, \mathcal{M}_f^c)$}
    \label{alg:fi}
    \renewcommand{\algorithmicrequire}{\textbf{Input:}}
    \renewcommand{\algorithmicensure}{\textbf{Output:}}
    
    \begin{algorithmic}[1]
        \REQUIRE (1) clean EEG $x$, trigger EEG $x_c^t$ from class $c$, interpolating ratio $\alpha$, \\
        (2) learned strategies $\mathcal{M}_e^c$, $\mathcal{M}_f^c$.
        \ENSURE the poisoned EEG $x^p$.
        
        \STATE $\mathcal{M}^c \leftarrow$ a zero matrix with the shape of $E \times F$

        \FOR{each $i \in \mathcal{M}_e^c$}
            \FOR{each $j \in \mathcal{M}_f^c$}
                \STATE $\mathcal{M}^c[i,j] \leftarrow 1$
            \ENDFOR
        \ENDFOR
        
        

        \STATE $\mathcal{A}_x, \mathcal{P}_x, \mathcal{A}_{x_c^t} \leftarrow \mathcal{F}^A(x), \mathcal{F}^P(x)$, $ \mathcal{F}^A(x_c^t)$

        \STATE $\mathcal{A}_x^P \leftarrow [(1-\alpha)\mathcal{A}_x + \alpha \mathcal{A}_{x_c^t}] \odot \mathcal{M}^{c}$ $+ \mathcal{A}_{x}\odot(1-\mathcal{M}^{c})$

        \STATE $x^p \leftarrow \mathcal{F}^{-1}(\mathcal{A}_x^P, \mathcal{P}_x)$
        
        \RETURN $x^p$
    \end{algorithmic}
\end{algorithm}
\end{minipage}
\end{wrapfigure}
After selecting the \textit{C} triggers from each class and learning the strategy for each trigger, the poisoned data are generated by injecting these triggers into clean data with the corresponding strategies.
As shown in Fig~\ref{fig:manibci}(c), given a clean data $x_i \in \mathcal{D}_{p}$ with label $c_i$, and a trigger data $x_{c_i}^t$,  let $\mathcal{F}^A$ and $\mathcal{F}^P$ be the amplitude and phase components of the fast Fourier
transform (FFT) result of a EEG signals, we denote the amplitude and phase spectrum of $x_i$ and $x_{c_i}^t$ as:
\begin{equation}
\begin{aligned}
    \mathcal{A}_{x_i} = \mathcal{F}^A(x_i), \mathcal{A}_{x_{c_i}^t} = \mathcal{F}^A(x_{c_i}^t), \\
    \mathcal{P}_{x_i} = \mathcal{F}^P(x_i), \mathcal{P}_{x_{c_i}^t} =\mathcal{F}^P(x_{c_i}^t).
\end{aligned}
\end{equation}
The new poisoned amplitude spectrum $\mathcal{A}_{x_i}^P$ is produced by linearly interpolating $\mathcal{A}_{x_i}$ and $\mathcal{A}_{x_{c_i}^t}$.
In order to achieve this, we produce a binary mask $\mathcal{M}^{c_i} \in \mathbb{R}^{E\times F}  = 1_{(j,k)}, j\in \mathcal{M}_e^{c_i}, k\in \mathcal{M}_f^{c_i} $, whose value is 1 for all positions corresponding to elements in both electrode and frequency strategies and 0 elsewhere.
Denoting $\alpha \in (0, 1]$ as the linear interpolating ratio, the new poisoned amplitude spectrum can be computed as follows, where $\odot$ indicates Hadamard product:
\begin{equation}
\begin{aligned}
    \mathcal{A}_{x_i}^P = \ &[(1-\alpha)\mathcal{A}_{x_i} + \alpha\mathcal{A}_{x_{c_i}^t}] \odot \mathcal{M}^{c_i} + \mathcal{A}_{x_i}\odot(1-\mathcal{M}^{c_i}).
\end{aligned}
\end{equation}
Finally, we adopt the injected poisoned amplitude spectrum $\mathcal{A}_{x_i}^P$ and the clean phase spectrum $\mathcal{P}_{x_i}$ to get the poisoned data by inverse FFT $\mathcal{F}^{-1}$:
    $x_i^p = \mathcal{F}^{-1}(\mathcal{A}_{x_i}^P, \mathcal{P}_{x_i}).$
The detailed procedure is written in Algorithm~\ref{alg:fi}.
By generating $x_i^p$ through this frequency injection approach, we obtain a subset $\mathcal{S}_p = \{x_1^p, ..., x_M^p\}$, which will combine with $\mathcal{D}_{train}$ to form the whole traing dataset $\mathcal{S}$. The EEG BCI model $f$ is then trained with $\mathcal{S}$ to obtain the ability of behvaing as equation~\ref{MTBA}.
\section{Experiment Settings}
\subsection{Datasets}
We demonstrate the effectiveness and generalizability of the proposed Professor X backdoor through comprehensive experiments on three EEG datasets. Some meta information is displayed in Table~\ref{tab:dataset}, where can be seen that these datasets vary significantly in tasks, electrode numbers, montages, and sampling rates. More details about preprocessing are illustrated in Appendix \ref{apd:dataset}. Our goal is to develope a task-agnostic and format-agnostic BA method for EEG BCI. Hence, these elaborately chosen datasets can effectively validate the generalizability of each BA method.

\vspace{-0.3cm}
\begin{table}[H]
  \setlength\tabcolsep{6pt}
  \caption{Meta information of the three datasets}
  \label{tab:dataset}
  \centering
  \begin{tabular}{lccccc}
    \toprule
    {Dataset} & { \# Class} & {\# Subject} & {\# Electrode} & Sampling Rate & Montage \\
    \midrule
    Emotion Recognition & 3 & 15 & 62 & 200 Hz & unipolar \\
    Motor Imagery & 4 & 9 & 22 & 250 Hz & unipolar \\
    Epilepsy Detection & 4 & 23 & 23 & 256 Hz & bipolar \\
    \bottomrule
  \end{tabular}
\end{table}
\vspace{-0.3cm}

\textbf{Emotion Recognition (ER) Dataset.} SEED \citep{zheng2015investigating} is a discrete EEG emotion dataset studying three types of emotions: happy, neutral, and sad.
SEED collected EEG from 15 subjects.
The EEG is recorded while the subjects were watching emotional videos.

\textbf{Motor Imagery (MI) Dataset.} BCIC-IV-2a \citep{brunner2008bci} dataset recorded EEG from 9 subjects while they were instructed to imagine four types of movements: left hand, right hand, feet, and tongue.

\textbf{Epilepsy Detection (ED) Dataset.} CHB-MIT \citep{shoeb2010application} is an epilepsy dataset required from 23 patients.
We cropped and resampled the CHB-MIT dataset to build an ED dataset with four types of EEG: ictal, preictal, postictal, and interictal phase EEG.

\subsection{Baselines}
\textbf{Non-stealthy Baselines.}
As mentioned in previous sections, to the best of our knowledge, Professor X is the first work that studies multi-trigger and multi-target class (MT) backdoor in EEG BCI.
For comparison, we design several baseline approaches which can be divided into two main groups: non-stealthy and stealthy.
Non-stealthy attacks contains \textbf{PatchMT} and \textbf{PulseMT}.
For a benign EEG segment $x \in \mathbb{R} ^ {E\times T}$.
PatchMT is a multi-trigger and MT extension of BadNets \citep{gu2019badnets} where we fill the first $\beta T$ timepoints of a EEG segments with a constant number, \textit{e.g.}, \{0.1, 0.3, 0.5\} for three-class task.
PulseMT is a multi-trigger and MT extension of NPP-based backdoor attacks \citep{meng2023eeg} where we use NPP signals with different amplitudes, \textit{e.g.}, \{-0.8, -0.3, 0.3, 0.8\} for different target classes.

\textbf{Stealthy Baselines.}
Previous works generate stealthy poisioned samples by controlling the training stage and can only attack single target class \citep{meng2024adversarial, ding2022towards, jiang2023backdoor}.
As they control the training of target model, it is unfair to directly compare their methods with Professor X.
There is no stealthy MT BA for EEG.
Thus, we design two MT stealthy attacks baselines: \textbf{CompMT} and \textbf{AdverMT}.
CompMT generates poisoned samples for different target classes by compressing the amplitude of EEG with different ratios, \textit{e.g.}, \{-0.1, 0, 0.1\} for three-class task.
AdverseMT is a multi-trigger and MT extension of adversarial filtering based attacks \citep{meng2024adversarial}, where we using a local model trained only on $\mathcal{S}_p$ to generate different spatial filters $\textbf{W}^*_i$ for different target classes, then we apply these spatial filters to generate poisoned samples. More details are written in Appendix \ref{apd:baselines}.

\subsection{Experimental Setup}
We follow the poisoning attack setting as the previous works \citep{meng2023eeg} and consider three widely-used EEG BCIs for classifier $f$: EEGNet \citep{lawhern2018eegnet}, DeepCNN \citep{schirrmeister2017deep}, and LSTM \citep{tsiouris2018long}.
We use a cross-validation setting to evaluate all BAs, each EEG dataset $\mathcal{D}$ is divided into three parts: training set $\mathcal{D}_{train}$, poisoning set $\mathcal{D}_p$, and test set $\mathcal{D}_{test}$.
Specifically, for a dataset contains $n$ subjects,
we select one subject's data as $\mathcal{D}_p$ one by one, and the remaining $n-1$ subjects to perform leave-one-subject-out (LOSO) cross-validation, \textit{i.e.}, one of the subjects as $\mathcal{D}_{test}$, and the remaining $n-2$ subjects as $\mathcal{D}_{train}$ (one of the subjects in $\mathcal{D}_{train}$ is chosen to be validation set).
In summary, for a dataset contains $n$ subjects, there are $n(n-1)$ runs to validate each EEG BCI backdoor attack method.
A poisoned subset $\mathcal{S}_p$ of $M$ ($M<N$) examples is generated based on $\mathcal{D}_p$.
Then $\mathcal{S}_p$ is combined with $\mathcal{D}_{train}$ to acquire $\mathcal{S} = \{\mathcal{S}_p, \mathcal{D}_{train}\}$. The poisoning ratio is defined as : $\rho = M/N$.

For all methods, we train the classifiers using the Adam optimizer with learning rate of 0.001. The batch size is 32 and the number of epochs is 100.
For all datasets and baselines, the interpolating ratio $\alpha=$ 0.8, the frequency poisoning ratio $\beta = $ 0.1, the electrode poisoning ratio $\gamma=$ 0.5.
For the reinforcement learning, we train $\pi_{\theta}$ networks $K=250$ epochs using the Adam optimizer with learning rate of 0.01.
The hyperparameters in advantage function is set to $\lambda=2$, $\mu = 0.3$, and $\nu = 0.005$.
More details of the experimental setup can be found in Appendix~\ref{apd:baselines}.

\section{Experimental Results}
\subsection{Effectiveness of Professor X}
This section presents the attack success rates of Professor X and baselines.
To evaluate the performance in the multi-trigger multi-payload scenario, for each test sample $(x, y) \in \mathcal{D}_{test}$, we enumerate all possible target labels $c_i \in \mathcal{C}$ including the true label $y$ and inject the trigger to activate the backdoor.
The attack is successful only when the backdoor classifier $f$ correctly predicts $c_i$ for each poisoned input $x$ with a target label $c_i$.

\subsubsection{Attack Performance}
The CA (Clean) and ASR (Attack) for each class of all attack methods on three EEG tasks with three EEG BCI models are presented in Table~\ref{tab:attack}.
The AdverMT, designed for single-target attack, fails to attacks multiple target classes.
While PulseMT achieves the second best on ER and ED dataset, CompMT achieves the second best on the MI dataset, indicating that these baselines are less generalizable.
Our Professor X significantly outperforms baselines at almost all cases ($p<0.05$) except attacking DeepCNN on the ED dataset, having ASRs above 0.8 on three datasets and even achieving an ASR of 1.000 on the MI dataset.
These results demonstrate that our Professor X is effective across different EEG tasks and EEG models, showcasing it's generalizability.

\begin{table}
  \setlength\tabcolsep{2pt}
  \caption{The clean accuraciy and attack success rate for each target class with 40\% poisoning rate. The best results are in \textbf{bold} and the second best are \underline{{underlined}}.}
  \label{tab:attack}
  \centering
  \scriptsize{
  \centering
  \begin{tabular}{l|l|ccccc|cccccc|cccccc}
    \toprule
    ~ & Dataset & \multicolumn{5}{c}{Emotion Recognition} & \multicolumn{6}{c}{Motor Imagery} & \multicolumn{6}{c}{Epilepsy Detection} \\
    \cmidrule(r){3-7} \cmidrule(r){8-13} \cmidrule(r){14-19}
     ~ & Method &  Clean & ASR & 0  & 1  & 2 & Clean & ASR & 0  & 1  & 2 & 3 & Clean & ASR & 0 & 1 & 2 & 3\\
    \midrule
    \multirow{6}*{\rotatebox{90}{EEGNet}} & No Attack & 0.477 & 0.333 & - & - & - & 0.327 & 0.250 & - & - & - & -  & 0.508 & 0.250 & - & - & - & -  \\
     ~ & PatchMT & \underline{0.492}  & 0.382 & 0.577 & 0.232  & 0.337  & {\underline{0.283}} & 0.824 & 0.866 & 0.880  & 0.787 & 0.762 & {\underline{0.460}} & {0.549} & {0.532} & {0.430} & 0.388 & 0.845  \\
    ~ & PulseMT & 0.463  & \underline{0.778} & \textbf{0.844} & \underline{0.509} & \textbf{0.981} & 0.270  & 0.825 & {\underline{0.947}} & 0.656 & 0.758 & 0.938 & 0.439 & \underline{0.810} & \underline{0.853} & \underline{0.745} & \underline{0.729} & {0.913} \\
    ~ & CompMT & 0.443  & 0.385 & 0.099 & 0.377 &0.678 & 0.269  & {\underline{0.865}} & 0.530 & \underline{0.997} & {\underline{0.983}}& {\underline{0.948}}  & 0.437 & 0.547 & 0.261 & 0.280 & 0.714 & \underline{0.933}  \\
    ~ & AdverMT & 0.457 & 0.334 & 0.276 & 0.330 & 0.396 & 0.257 & 0.243 & 0.316 & 0.192 & 0.230 & 0.235 & 0.413 & 0.250 & 0.326 & 0.264 & 0.200 & 0.210 \\
    ~ & Professor X & \textbf{0.535}  & \textbf{0.857} & \underline{0.831} & \textbf{0.791}  & \underline{0.949} & \textbf{0.323} & \textbf{1.000} & \textbf{0.999} & \textbf{1.000} & \textbf{1.000} & \textbf{0.999} & \textbf{0.477} & \textbf{0.944} &\textbf{ 0.930} & \textbf{0.954} & \textbf{0.921} & \textbf{0.970} \\
    \midrule
     \multirow{6}*{\rotatebox{90}{DeepCNN}} & No Attack & 0.497 & 0.333 & - & - & - & 0.301 & 0.250 & - & - & - & -  & 0.443 & 0.250 & - & - & - & -  \\
     ~ & PatchMT & \underline{0.481}  & 0.342 & 0.248 & 0.323  & 0.453  & 0.276 & 0.704 & 0.638 & 0.977  & 0.774 & 0.425 & 0.431 & 0.729 & 0.416 & \textbf{0.890} & 0.719 & 0.892  \\
    ~ & PulseMT & 0.450  & \underline{0.596} & \textbf{0.815} & 0.334  & \underline{0.638}  & 0.261 & 0.829 & {\underline{0.764}} & 0.968  & 0.819 & 0.765 & 0.405 & \textbf{0.885} & \textbf{0.872} & {\underline{0.862}} & \textbf{0.861} & \textbf{0.943}  \\
    ~ & CompMT & 0.461  & 0.427 & 0.473 & \underline{0.473}  & 0.336  & {\underline{0.286}} & {\underline{0.887}} & 0.638 & {\underline{0.982}}  & {\underline{0.946}} & {\underline{0.980}} & {\underline{0.446}} & 0.538 & 0.196 & 0.466 & 0.571 & {\underline{0.918}}  \\
    ~ & AdverMT & 0.367  & 0.388 & 0.298 & 0.453  & 0.412  & 0.245 & 0.247 & 0.320 & 0.221  & 0.196 & 0.240 & 0.396 & 0.275 & 0.354 & 0.218 & 0.227 & 0.301  \\
    ~ & Professor X & \textbf{0.534}  & \textbf{0.832} & \underline{0.732} & \textbf{0.865}  & \textbf{0.901}  & \textbf{0.315} & \textbf{1.000} & \textbf{1.000} & \textbf{1.000}  & \textbf{1.000} & \textbf{0.999} & \textbf{0.469} & {\underline{0.828}} & {\underline{0.725}} & 0.839 & {\underline{0.845}} & 0.904   \\
    \midrule
     \multirow{6}*{\rotatebox{90}{LSTM}} & No Attack & 0.506 & 0.333 & - & - & - & 0.264 & 0.250 & - & - & - & -  & 0.462 & 0.250 & - & - & - & -  \\
     ~ & PatchMT & 0.509  & 0.368 & 0.311 & 0.392  & 0.401  & 0.261 & 0.429 & 0.395 & 0.296  & 0.386 & 0.639 & 0.450 & 0.513 & 0.500 & 0.437 & 0.417 & 0.700   \\
    ~ & PulseMT & \underline{0.511}  & \underline{0.824} & \underline{0.883} & \underline{0.645}  & \underline{0.943}  & \textbf{0.265} & 0.533 & {\underline{0.787}} & 0.327  & 0.282 & 0.737 & {\underline{0.451}} & {\underline{0.804}} & \textbf{0.845} & {\underline{0.769}} & {\underline{0.709}} & {\underline{0.895}}  \\
    ~ & CompMT & 0.484  & 0.490 & 0.272 & 0.269  & 0.929  & 0.260 & {\underline{0.548}} & 0.219 & {\underline{0.511}}  & {\underline{0.523}} & {\underline{0.940}} & \textbf{0.455} & 0.435 & 0.194 & 0.217 & 0.490 & 0.840   \\
    ~ & AdverMT & 0.367  & 0.415 & 0.472 & 0.453  & 0.321  & 0.239 & 0.271 & 0.308 & 0.215  & 0.247 & 0.312 & 0.432 & 0.268 & 0.367 & 0.232 & 0.198 & 0.275   \\
    ~ & Professor X & \textbf{0.519}  & \textbf{0.954} & \textbf{0.998} & \textbf{0.868}  & \textbf{0.996}  & {\underline{0.264}} & \textbf{0.966} & \textbf{0.987} & \textbf{0.988}  & \textbf{0.901} & \textbf{0.986} & 0.444 & \textbf{0.865} & {\underline{0.795}} &\textbf{ 0.833} & \textbf{0.857} & \textbf{0.975}   \\
    \bottomrule
  \end{tabular}
  }
\end{table}

\subsubsection{Performance of the Reinforcement Learning: Policy Gradient}
Displaying in Table \ref{tab:rl}, the performance of the policy gradient was compared with other common optimazation algorithms, including genetic algorithm (GA) \citep{katoch2021review} and random selection (The search space is too large for performing grid search as explained in Section \ref{sec:rl}).
It can be observed that the policy gradient outperforms GA while only spending $16\%$ training time of GA.
We plot the learning curve of RL in Appendix \ref{apd:rl_vis}, which demonstrates that RL learns well strategies within 50 epochs, i.e., only trains 50 backdoor models and saves lots of time.
It is worth mentioning that the random algorithm achieves not bad results, proving that our methods can be applied without RL if some performance drop is acceptable.

\vspace{-0.4cm}
\begin{table}[H]
  \setlength\tabcolsep{5pt}
  \caption{Clean and attack performance with with different trigger
search optimization algorithms, the poisoning rate is set to 10\%. The target model is EEGNet.}
  \label{tab:rl}
  \centering
  \small{
  \begin{tabular}{llccccccccc}
    \toprule
    \multirow{2}*{\diagbox{Method}{Dataset}} & \multicolumn{3}{c}{Emotion} & \multicolumn{3}{c}{Motor Imagery} & \multicolumn{3}{c}{Epilepsy} \\
    \cmidrule(r){2-4} \cmidrule(r){5-7} \cmidrule(r){8-10}
     ~ &  Clean  & Attack & Time $\downarrow$ &  Clean  & Attack & Time $\downarrow$&  Clean  & Attack & Time $\downarrow$\\
    \midrule
    Random  & 0.520  & 0.771 & - & 0.291 & 0.857 & -  & 0.501 & 0.721 & - \\
    Genetic Algorithm & 0.516 & 0.826& 15.2h  & 0.302 & 1.000 & 10.0h  & 0.492 & 0.862 & 30.5h  \\
    Policy Gradient & 0.535 & 0.857& 2.5h  & 0.323 & 1.000 & 1.8h  & 0.477 & 0.944 & 5.2h  \\
    \bottomrule
  \end{tabular}
  }
\end{table}

\subsubsection{Performance of Learned Mask Strategies on Other Target Models}
We demonstrate that the injecting strategies learned on a EEG classifier $f$ can be used to attack other EEG classifiers $\hat{f}$.
In other words, Marksman can still be effective when the adversary has no knowledge of the target models $\hat{f}$. To perform the experiments, we use the strategy learned with a classifier $f$, then generate poisoned samples to attack another classifier $\hat{f}$ whose network is different from $f$. Table \ref{tab:diff_RL} shows the performance difference, it can be observed that the difference is relatively small in most of the cases, demonstrating the transferability of the injecting strategy learned with reinforcement learning.
\vspace{-0.4cm}
\begin{table}[H]
  \setlength\tabcolsep{4pt}
  \caption{Clean and attack performance on other models. Red values represent the decreasing performance in attacks with $f$ is the same as $\hat{f}$. Blue values mean increments or unchanged .}
  \label{tab:diff_RL}
  \centering
  \small{
  \begin{tabular}{lcccccccccccc}
    \toprule
     \multirow{2}*{Models} & \multicolumn{4}{c}{$f$ : EEGNet} & \multicolumn{4}{c}{$f$ : DeepCNN} & \multicolumn{4}{c}{$f$ : LSTM} \\
    \cmidrule(r){2-5} \cmidrule(r){6-9} \cmidrule(r){10-13}
     ~ &  \multicolumn{2}{c}{$\hat{f}$ : DeepCNN}  & \multicolumn{2}{c}{$\hat{f}$ : LSTM}  &  \multicolumn{2}{c}{$\hat{f}$ : EEGNet}  & \multicolumn{2}{c}{$\hat{f}$ : LSTM} &  \multicolumn{2}{c}{$\hat{f}$ : EEGNet}  & \multicolumn{2}{c}{$\hat{f}$ : DeepCNN} \\
     \cmidrule(r){2-3} \cmidrule(r){4-5} \cmidrule(r){6-7} \cmidrule(r){8-9} \cmidrule(r){10-11} \cmidrule(r){12-13}
     Datasets &  Clean  & Attack  &  Clean  & Attack &  Clean  & Attack &  Clean  & Attack  &  Clean  & Attack &  Clean  & Attack \\
    \midrule
    \multirow{2}*{Emotion} & 0.458  & 0.781 & 0.485 & 0.938  & 0.516 & 0.813 & 0.490 & 0.936  & 0.516 & 0.863 & 0.497 & 0.779  \\
    ~ & \textcolor{red}{0.026}  & \textcolor{red}{0.051} & \textcolor{red}{0.034} & \textcolor{red}{0.016} & \textcolor{red}{0.019}  & \textcolor{red}{0.044} & \textcolor{red}{0.029} & \textcolor{red}{0.018}& \textcolor{red}{0.019}  & \textcolor{blue}{0.006} & \textcolor{red}{0.037} & \textcolor{red}{0.053} \\
     \midrule
    \multirow{2}*{Motor} & 0.316  & 1.000 & 0.265 & 0.946  & 0.309 & 1.000 & 0.264 & 0.972  & 0.306 & 1.000 & 0.306 & 1.000  \\
    ~ & \textcolor{blue}{0.001}  & \textcolor{blue}{0.000} & \textcolor{blue}{0.001} & \textcolor{red}{0.020} & \textcolor{red}{0.014}  & \textcolor{blue}{0.000} & \textcolor{blue}{0.000} & \textcolor{blue}{0.006}& \textcolor{red}{0.017}  & \textcolor{blue}{0.000} & \textcolor{red}{0.009} & \textcolor{blue}{0.000} \\
     \midrule
    \multirow{2}*{Epilepsy} & 0.442  & 0.759 & 0.469 & 0.806  & 0.448 & 0.943 & 0.445 & 0.813  & 0.448 & 0.926 & 0.427 & 0.850  \\
    ~ & \textcolor{red}{0.027}  & \textcolor{red}{0.069} & \textcolor{blue}{0.025} & \textcolor{red}{0.059} & \textcolor{red}{0.029}  & \textcolor{red}{0.001} & \textcolor{blue}{0.001} & \textcolor{red}{0.052}& \textcolor{red}{0.029}  & \textcolor{red}{0.018} & \textcolor{red}{0.042} & \textcolor{blue}{0.022} \\
    \bottomrule
  \end{tabular}
  }
\end{table}

\subsubsection{Attack Performance with Different Hyperparameters}
\begin{figure}[H]
    \centering
    \setlength{\abovecaptionskip}{-0.3cm}
    \includegraphics[width=1.0\textwidth]{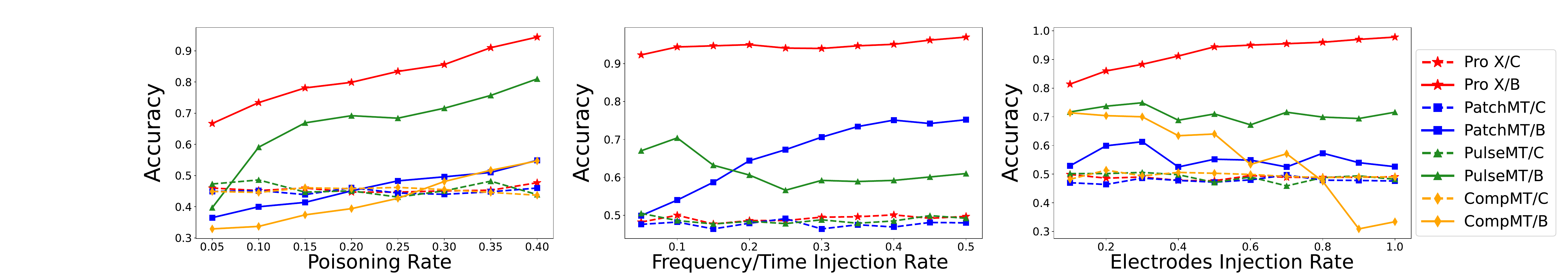}
    \caption{Clean (/C) and attack (/B) performance with different poisoning or injection rates.}
    \label{fig:hyperpara}
\end{figure}

We investigate the influences of three different hyperparameters: poisoning rate $\rho$, frequency injection rate $\beta$, and electrode injection rate $\gamma$.
The performance of attacking EEGNet on the ED dataset are displayed in Fig \ref{fig:hyperpara}. It can be seen that the ASRs are positively correlated with poisoning rate.
Note that it is non-trivial for multi-target class attack, thus the ASR is not high compared to the single class attack.
Professor X outperforms other attacks in all cases and is robust to the change of $\beta$ and $\gamma$.

\subsection{Robustness of Professor X}
In this section, we evaluate the robustness of our Professor X against different EEG preprocessing method and various representative backdoor defenses.

\subsubsection{Robustness against EEG Preprocessing Methods}
To develop an EEG BCI, it is very common to preprocess the raw EEG signals, \textit{e.g.}, 1) band-stop filtering and 2) down-sampling.
An EEG backdoor attack is impractical in real scenarios if it is no longer effective when the target model is trained with the preprocessed poisoned EEG.
Hence, we must take the robustness against preprocessing methods into account, which is widely ignored in the image backdoor attack field.
The performance of each method facing different preprocessing methods are presented in Table \ref{tab:preprocess}.
It can be observed that our Professor X is robust in all cases.
However, when removing the DIS loss, the performance of Professor X decreases a lot after EEG preprocessing, especially facing the 30 Hz high-stop filtering preprocessing due to the HF loss that encourages the policy network learns to injecting high frequency.

\begin{table}[H]
  \vspace{-0.4cm}
  \setlength\tabcolsep{6pt}
  \caption{Clean and attack performance on three datasets after different EEG preprocessing methods. The target model is EEGNet. M w.o. DIS means removing the DIS loss in Professor X.}
  \label{tab:preprocess}
  \centering
  \small{
  \begin{tabular}{l|lccccccccc}
    \toprule
    \multirow{2}*{\makecell[l]{}} & Preprocessing & \multicolumn{2}{c}{No defense}& \multicolumn{2}{c}{20 Hz low} & \multicolumn{2}{c}{30 Hz high} & \multicolumn{2}{c}{25\% down} & Average\\
    \cmidrule(r){3-4} \cmidrule(r){5-6} \cmidrule(r){7-8} \cmidrule(r){9-10}
     ~ & Method &  Clean  & Attack  &  Clean  & Attack &  Clean  & Attack &  Clean  & Attack & ASR\\
    \midrule
    \multirow{2}*{\rotatebox{90}{ER}} & Professor X & 0.535   & 0.857  & 0.512  & 0.829   & 0.463  & 0.892   & 0.518  & 0.908  & 0.876 \\
    ~ & w/o DIS & 0.506   & 0.859  & 0.492  & 0.816   & 0.466  & 0.333 & 0.498  & 0.807  & 0.652   \\
    \midrule
    \multirow{2}*{\rotatebox{90}{MI}} & Professor X & 0.323   & 1.000  & 0.285  & 1.000   & 0.329  & 1.000   & 0.321  & 1.000  & 1.000 \\
    ~ & w/o DIS & 0.298   & 1.000  & 0.264  & 1.000   & 0.322  & 0.250 & 0.284  & 0.990  & 0.746   \\
    \midrule
    \multirow{2}*{\rotatebox{90}{ED}} & Professor X & 0.497   & 0.944  & 0.492  & 0.914   & 0.494  & 0.856   & 0.516  & 0.818  & 0.920 \\
    ~ & w/o DIS & 0.515   & 0.250  & 0.477  & 0.864   & 0.508  & 0.250 & 0.510  & 0.249  & 0.454 \\
    \bottomrule
  \end{tabular}
  }
  \vspace{-0.2cm}
\end{table}

\subsubsection{Robustness against Neural Cleanse: Trigger Inversion}
\begin{wrapfigure}[9]{r}{0.35\textwidth}
    \vspace{-0.65cm}
    \begin{center}
        \setlength{\abovecaptionskip}{-0.4cm}
        \includegraphics[width=0.35\textwidth]{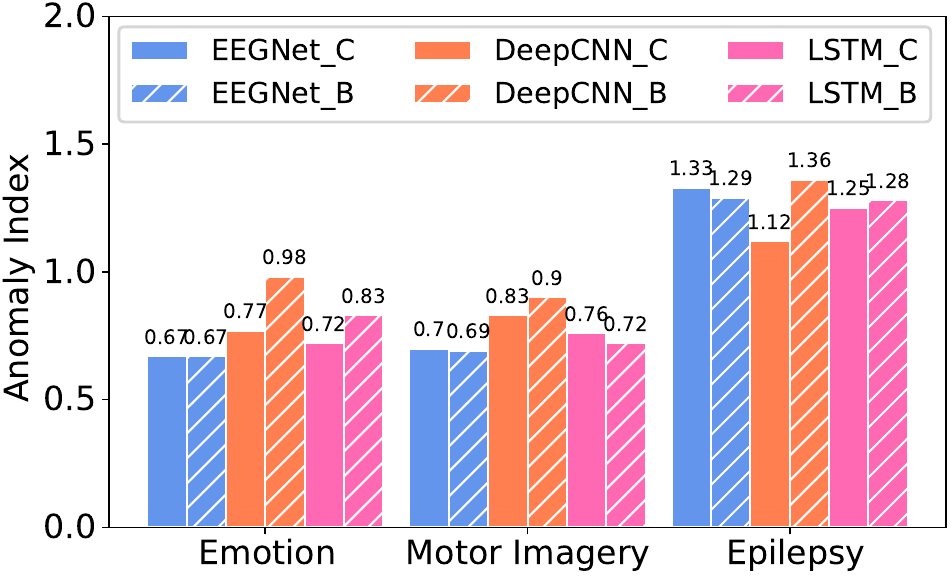}
        \caption{Anomaly Index of three models on three datasets.}
        \label{fig:ncbar}
    \end{center}
\end{wrapfigure}
Neural Cleanse (NC) \citep{wang2019neural} calculate a metric called Anomaly Index by reconstructing trigger pattern for each possible label.
The Anomaly Index is positively correlated with the size of the reconstruction trigger.
A model with Anomaly Index $>$ 2 is considered to be backdoor-injected.
We display the Anomaly Indexes of the clean models and the backdoor-injected model by Professor X in Fig \ref{fig:ncbar}. It can be seen that Professor X can easily bypass NC. The reconstructed trigger patterns on three datasets are presented in Appendix \ref{apd:nc}.

\vspace{0.1cm}
\subsubsection{Robustness against STRIP: Input Perturbation}
\begin{figure}[H]
    \vspace{-0.2cm}
    \centering
    \setlength{\abovecaptionskip}{-0.3cm}
    \includegraphics[width=1.0\textwidth]{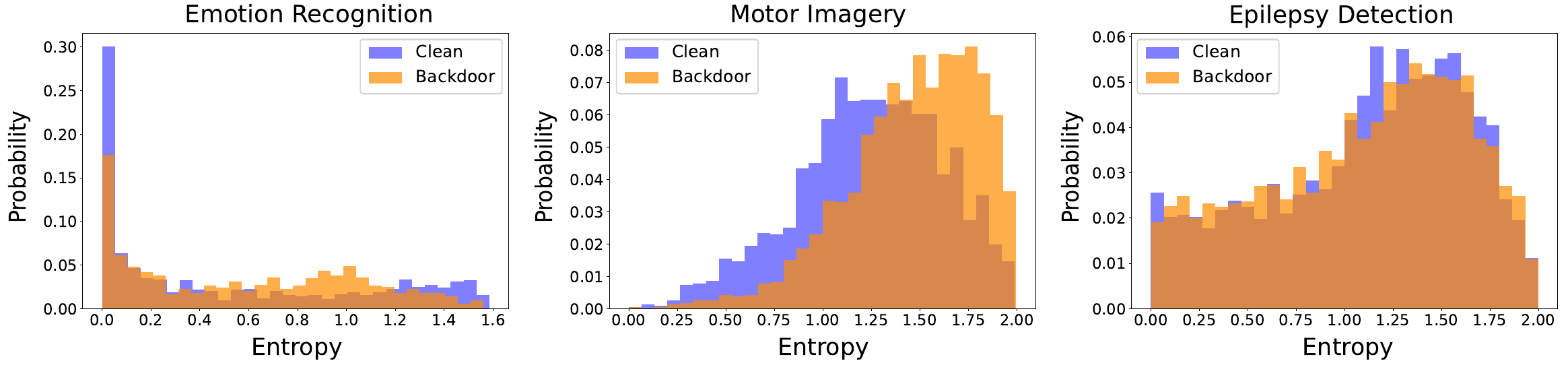}
    \caption{Performance against STRIP on three datasets, the target model is EEGNet.}
    \label{fig:strip}
\end{figure}

We evaluate the robustness of Professor X against STRIP \citep{gao2019strip}, which perturbs the input EEG and calculates the entropy of the predictions of these perturbed EEG data. Based on the assumption that the trigger is still effective after perturbation, the entropy of backdoor input tends to be lower than that of the clean one. The results are plotted in Fig \ref{fig:strip}, it can be seen that the entropy distributions of the backdoor and clean samples are similar.

\vspace{-0.4cm}

\subsubsection{Robustness against Spectral Signature: Latent Space Correlation}
\begin{figure}[H]
    \centering
    \setlength{\abovecaptionskip}{-0.3cm}
    \includegraphics[width=1.0\textwidth]{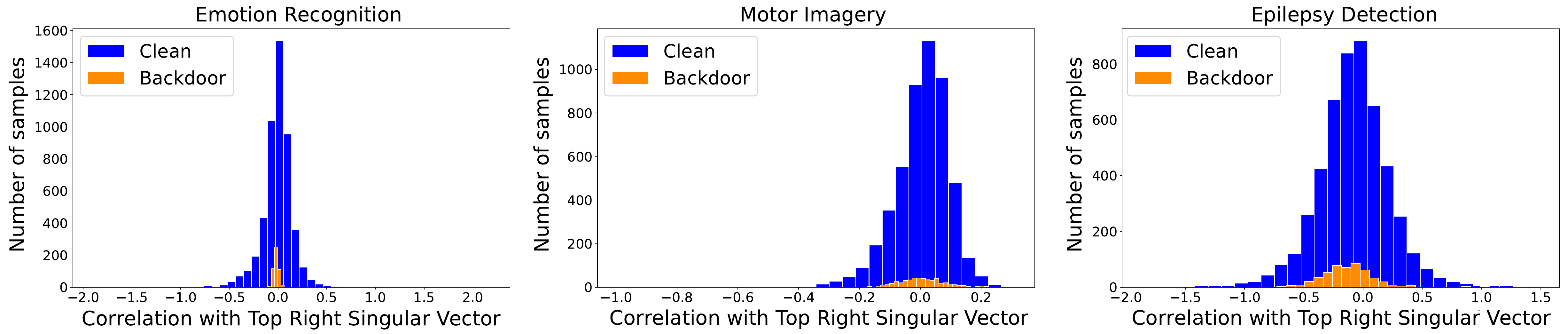}
    \caption{Performance against Spectral Signature on three datasets, the target model is EEGNet.}
    \label{fig:spectral}
\end{figure}

Spectral Signature \citep{tran2018spectral} detects the backdoor samples by statistical analysis of clean data and backdoor data in the latent space.
Following the same experimental settings in \citep{tran2018spectral}, we randomly select 5,000 clean samples and 500 Professor X backdoor samples and plot the histograms of the correlation scores in Fig \ref{fig:spectral}.
There is no clear separation between these two sets of samples, showing the stealthiness of Professor X backdoor samples in the latent space.

\subsubsection{Robustness against Fine-Pruning}
\begin{wrapfigure}[9]{r}{0.35\textwidth}
    \vspace{-1.5cm}
    \begin{center}
        \setlength{\abovecaptionskip}{-0.3cm}
        \includegraphics[width=0.35\textwidth]{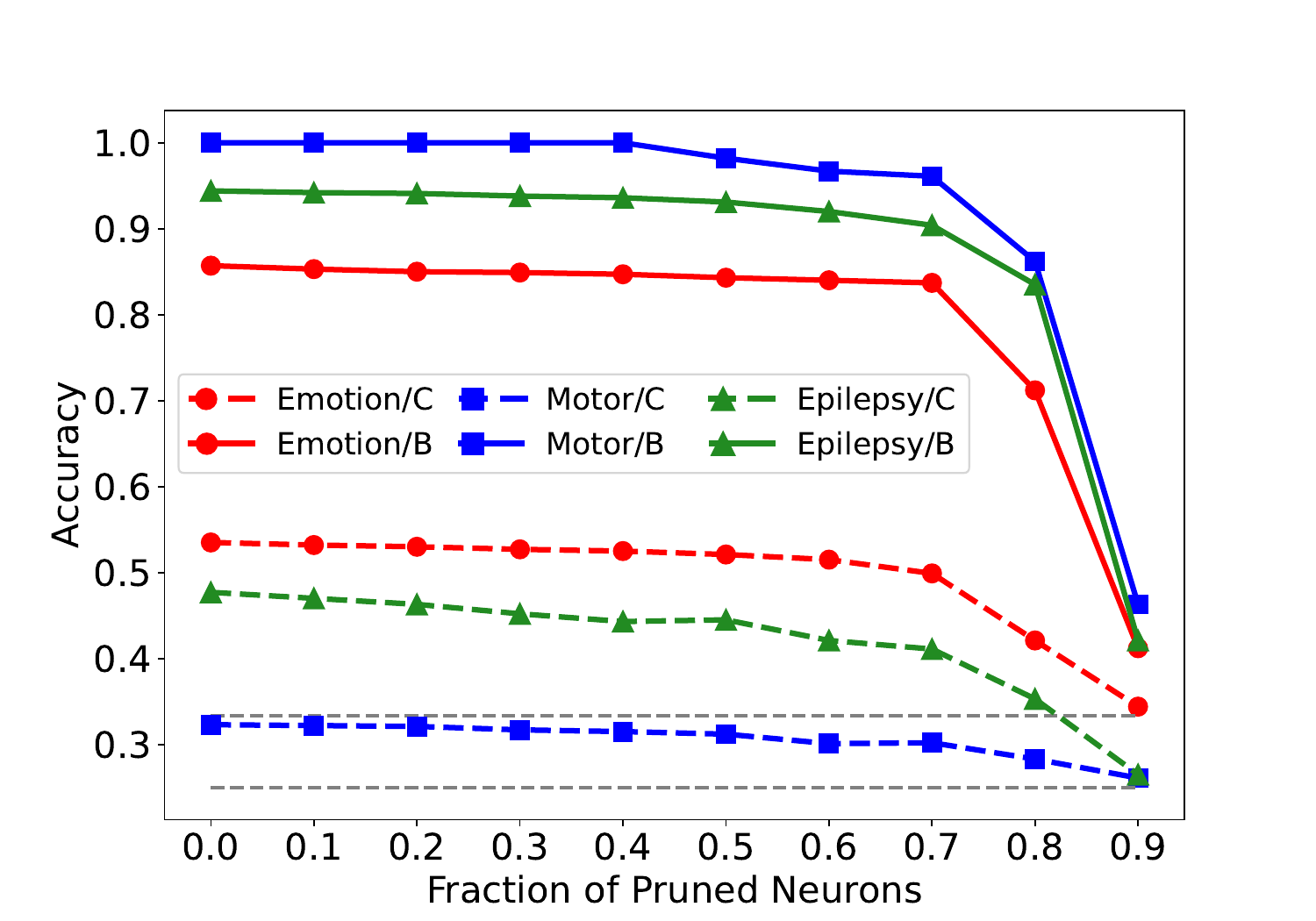}
        \caption{Performances of EEGNet against Fine-Pruning on three datasets.}
        \label{fig:fp}
    \end{center}
\end{wrapfigure}
We evaluate the robustness of Professor X against Fine-Pruning \citep{liu2018fine}, a model analysis based defense which finds a classifier’s low-activated neurons given a small clean dataset. Then it gradually prunes these low-activated neurons to mitigate the backdoor without affecting the CA.
We can observe from Fig \ref{fig:fp} that the ASR drops considerably small when pruning ratio is less than 0.7, suggesting that the Fine-Pruning is ineffective against Professor X.

\subsection{Visualization of Backdoor Attack Samples}
\begin{figure}[H]
    \centering
    \setlength{\abovecaptionskip}{-0.3cm}
    \includegraphics[width=1.0\textwidth]{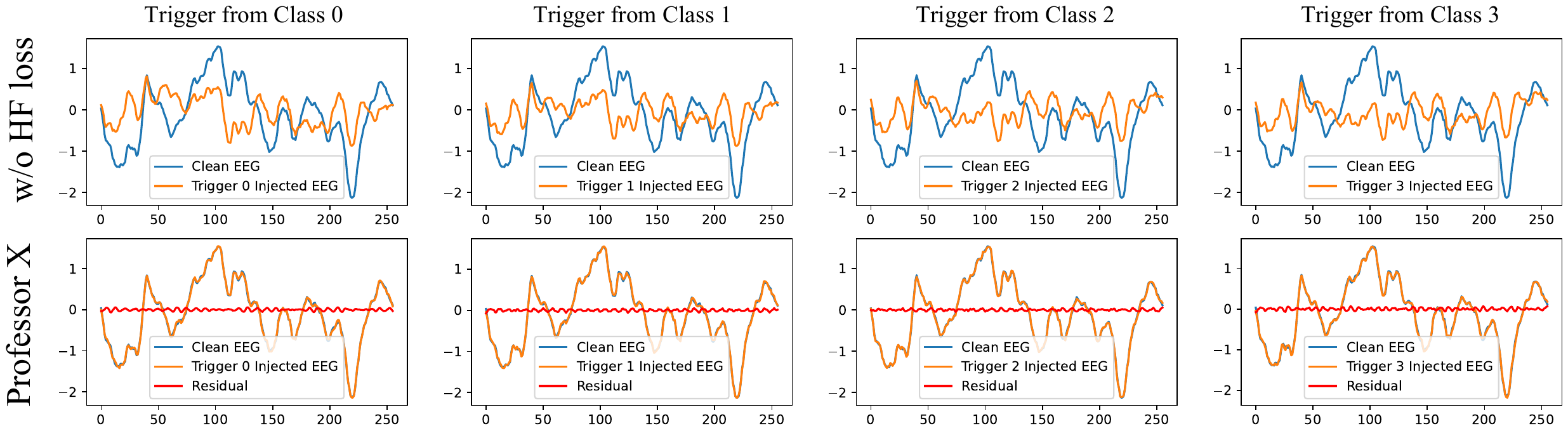}
    \caption{The Clean EEG (Blue), Trigger-injected EEG (Orange) and the Residual (Red) of the ED dataset. The \textit{x}-axis is the timepoints, the \textit{y}-axis is the normalized amplitude. Top row: w/o HF loss; Bottom row: with HF loss.
    Each column indicates each possible class.}
    \label{fig:vis_chb}
    \vspace{-0.3cm}
\end{figure}

To evade from human perception (\textbf{C2} in Section \ref{sec:rl}), we design to obatin injecting strategies with HF loss.
It can be seen from the bottom row of Fig \ref{fig:vis_chb} that Professor X (with HF loss) generates stealthy poisoned EEG, which is almost the same as the clean EEG, demonstrating the \textbf{High Stealthiness}.
The poisoned EEG will be conspicuous compared to the clean EEG if remove the HF loss.

\section{Conclusion}
In this paper, we proposed Professor X, a novel EEG backdoor for manipulating EEG BCI, where the adversary can arbitrarily control the output for any input samples.
To the best of our knowledge, Professor X is the first method that considers which EEG electrodes and frequencies to be injected for different EEG tasks and formats.
We specially design the reward function in RL to enhance the robustness and stealthiness.
Experimental results showcase the effectiveness, robustness, and generalizability of Professor X.
This work alerts the EEG community of the potential danger of the vulnerability of EEG BCI against BA and calls for defensive studies for EEG modality. It is worth noting that Professor X can also be applied for protecting  intellectual properties of EEG datasets and BCI models, offering a concealed and harmless approach to add authors' watermark (backdoor can be regarded as watermark), indicating the real-world application of Professor X.

\bibliographystyle{iclr2025_conference}
\bibliography{reference}

\clearpage
\appendix
\section{Limitations}
Our Professor X is a backdoor attack in the frequency domain, which requires to transform the EEG signals into frequency domain through fast Fourier transform (FFT) and return to temporal domain through inverse FFT (iFFT).
The operation of FFT and iFFT in the trigger injection function are a little more time-consuming compared to other backdoor attack directly in the temporal domain, like PatchMT \cite{gu2019badnets} and PulseMT \cite{meng2023eeg}. Future effort will be devoted into the faster implementation of FFT and iFFT, for example, taking the advantage of modern GPUs.

It is a little more time-consuming for the reinforcement learning to acquire the optimal strategies for each trigger. However, we can obtain a general injecting strategy for each EEG BCI tasks, which can achieve a relatively good performance without reinforcement learning, as we can see from Table \ref{tab:diff_RL} that random injection strategy has an acceptable performance.

\section{Broader Impacts}
With the rapid development of techniques, EEG BCIs gain a wide range of applications from health care to human-computer interaction.
Some companies like Neuralink adopt the EEG BCI to assist paralytic patients helping themselves in daily lives.
However, if the EEG BCI is backdoor attacked by Professor X, which allows the attacker to arbitrarily control BCI's outputs, the BCI users may fall into tremendous fatal troubles.
For instance, one paralytic patient controls his/her wheelchair by EEG BCI, the attacker can manipulate the wheelchair to run down a steep staircase.
For an epileptic patient, the attacker can let all the output be Normal State, even when the patient is experiencing an epileptic seizure.
This paper reveals the severe danger faced by EEG BCIs, demonstrating the possibility that someone can maliciously manipulate the outputs of EEG BCIs with arbitrary target class.

Professor X can also be used for positive purposes, like protecting intellectual property of EEG dataset and EEG models with watermarking. As our Professor X has a very small impact of the clean accuracy, and the poisoning approach is clean label poisoning, Professor X is a fantastic method for watermarking EEG dataset and models.

For a company that provides EEG dataset, it can select different EEG triggers for different customs to generate poisoned data and inject into the dataset provided to customs who buy the dataset. As a result, the company have the information of which trigger is corresponding to which customs, e.g., trigger \textit{x} is in the dataset provided to custom \textit{X}, trigger \textit{y} is in the dataset provided to custom \textit{Y}. If an EEG model from a company which didn't buy dataset is detected having this watermark (backdoor) with trigger \textit{x}, the company knows that the custom \textit{X} leaked the dataset.
Similarly, if an EEG model is detected having this watermark (backdoor) with trigger \textit{y}, the company knows that the custom \textit{Y} leaked the dataset.

\section{Datasets and Preprocessing}
\label{apd:dataset}

In this section, we introduce the three datasets used in our experiments, and explain the preprocessing. We elaborately selected these three datasets because of three reasons: \textbf{1)} They cover three different EEG tasks that are important and common in EEG BCI field; \textbf{2)} The EEG formats of these datasets vary significantly; \textbf{3)} The EEG tasks are all multi class classification tasks, that is, the number of categories is more than two. 
Experiments on these three datasets can validate the efficacy, manipulating performance, and generalizability of each BA methods as much as possible.

\subsection{Emotion Recognition (ER)}
The SJTU Emotion EEG Dataset (SEED) was incoporated as the representative dataset of emotion recogniton tasks \cite{zheng2015investigating}. It consists of EEG recordings from 15 subjects watching 15 emotional video clips with three repeated session each on different days. Each video clip is supposed to evoke one of the three target emotions: positive, neutral, and negative. The EEG signals were acquired by the 62-channel electrode cap at a sampling rate of 1000 Hz. We performed below preprocessing procedures for the 62-channel EEG signals: 1) Down-sampling from 1000 Hz to 200 Hz, 2) Band-pass filtering at 0.3-50 Hz, 3) Segmenting EEG signals into 1-second (200 timepoints), obtaining 3394 EEG segments in each session for each subject.

\subsection{Motor Imagery (MI)}
We employ the BCIC-IV-2a as a representative dataset of MI classification tasks \cite{brunner2008bci}. It contains EEG recordings in a four-class motor-imagery task from nine subjects with two repeated session each on different days. During the task, the subjects were instructed to imagine four types of movements (\textit{i.e.}, right hand, left hand, feet, and tongue) for four seconds. Each session consists of a total of 288 trials with 72 trials for each type of the motor imagery. The EEG signals were recorded by 22 Ag/AgCl EEG electrodes in a sampling rate of 250 Hz.
We segment the 22-channel EEG signals into 1-second segments, resulting in totally 1152 EEG data for each subject.

\subsection{Epilepsy Detection (ED)}
The CHB-MIT, one of the largest
and most used public datasets for epilepsy, is adopted as a representative dataset of ED tasks \cite{shoeb2010application}. It recorded 877.39 hours of multi-channel EEG in a sampling rate of 256 Hz from 23 pediatric patients with intractable seizures.
However, as the montages (\textit{i.e.}, the number and the places of electrodes) of EEG signals vary significantly among different subjects' recordings, we select to use only the EEG recordings with the same 23 channels (see Appendix A) and discard other channels or the recordings don't have all these 23 channels.
Due to the purpose is to test whether the backdoor attack works on the ED task, not to study the epilepsy EEG classification, we segment part of the CHB-MIT dataset to form a four-class ED dataset (\textit{i.e.}, the preictal, ictal, postictal, and interictal phases).
Specifically, for a ictal phase EEG recording of $t_i$ seconds from $[s_i, e_i]$ timepoints, we segment the $[s_i-t_i, e_i]$ EEG as the preictal phase, the $[e_i, e_i+t_i]$ EEG as the postictal phase, and another $t_i$ seconds EEG recordings as the interictal phase which satisfying there is no ictal phase within half an hour before or after.
Then we segment the 23-channel EEG signals into 1-second segments,
consequently, there are 41336 segments left in total from all subjects, 10334 for each phase. As the imbalanced amount of data across different subjects, we separate these 41336 segments into 10 groups and treat the ten groups as 10 subjects.

\section{Implementation Details}
\label{apd:baselines}

\subsection{Experiment Computing Resources}
We use two servers for conducting our experiments. A server with one Nvidia Tesla V100 GPU is used for running reinforcement learning, the CUDA version is 12.3.
Another server with four Nvidia RTX 3090 GPUs is used for running the backdoor attacks, the CUDA version is 11.4.

\subsection{Details of Baseline Methods}
In our Professor X backdoor attacks, for an EEG segment $x_i \in \mathbb{R}^{E \times T}$, we modify the $\beta F$ frequency-points and $\gamma E$ electrodes of a EEG segments with a constant number.

There are four baseline methods in our study for multi-target backdoor attacks, two of them are non-stealthy attacks (\textbf{PatchMT} and \textbf{PulseMT}) and two are stealthy attacks (\textbf{CompressMT} and \textbf{AdverseMT}).
In order to achieve a fair comparison, we modify only first $\gamma E$ electrodes for all baseline attack methods.
For the non-stealthy attacks, which are all on the temporal domains, we modify $\beta T$ timepoints of EEG signals.
For the stealthy attacks, there is no constraint of the numbers of the modify timepoints as these attacks achieve stealthiness in another way.

For each baseline method, we try our best to find out the best performance, as demonstrated below. We promise that we did not maliciously lower the performances of the baseline methods.

\subsubsection{PatchMT}
PatchMT is a multi-trigger and MT extension of BadNets \cite{gu2019badnets} where we fill the first $\beta T$ timepoints and $\gamma E$ electrodes of a EEG segments with a constant number.
Specifically, for an EEG segment $x_i \in \mathbb{R}^{E \times T}$, we set the first $\gamma E$ electrodes and the first $\beta T$ timepoints of the EEG segment to a constant number. 
We normalize the EEG segment $x_i \in \mathbb{R}^{E \times T}$ to let $\textbf{x}_i$'s mean is 0 and std is 1. Then set the first $\gamma E$ electrodes and the first $\beta T$ timepoints of $\textbf{x}_i$ to a different constant number for different class. The constant number for each class of $\{0, 1, 2, 3\}$ for four classes, and \{-0.1, 0.0, 1.0\} for three classes. Finally, denormalize $\textbf{x}_i$ to original signal $x_i$'s scale to generate $x^p_i$.

Although we try our best to find the best performance of PatchMT, and BadNets \cite{gu2019badnets} is really efficient in image backdoor attacks, PatchMT cannot have satisfactory results in EEG BCI attack.

\subsubsection{PulseMT}
For PulseMT, we met the same questions as the PatchMT: how to identify the amplitude of each NPP signal for each class? If the numbers are too large then normal EEG signals, it will be unfair. If the numbers are too small, the efficacy of PulseMT is too negative.

We normalize the EEG segment $x_i \in \mathbb{R}^{E \times T}$ to let $\textbf{x}_i$'s mean is 0 and std is 1. The constant amplitude for each class of $\{-0.8, -0.3, 0.3, 0.8\}$. Finally, denormalize $\textbf{x}_i$ to original signal $x_i$'s scale to generate $x^p_i$.

\subsubsection{CompressMT}
Compressing the amplitude of EEG signals in the temporal domain will not change the morphology and the frequency distribution of EEG signals, thus obtaining stealthiness.
For three-class Emotion datasets, the compress rate is \{0.8, 0.6, 0.4\}.
For four-class Motor Imagery and Epilepsy datasets, the compress rate is \{0.8, 0.6, 0.4, 0.2\}.

\subsubsection{AdverseMT}
AdverseMT is another stealthy EEG backdoor attacks, which is the multi-trigger and multi-target extension of adversarial spatial filter attacks \cite{meng2024adversarial},
in wihch, for EEG segment $x_i \in \mathbb{R}^{E \times T}$, it learns an Spatial Filter $\textbf{W} \in \mathbb{R}^{E\times E}$ by the adversarial loss to let the model $f$ misclassify $x_i$:
\begin{align}
    \label{eq:adverse_loss}
    \mathop{\min}_{\textbf{W}} \mathbb{E}_{(x_i, y_i)\sim \mathcal{D}}
    [ - \mathcal{L}_{CE}(\textbf{W}x_i, y_i) + 
    \alpha \mathcal{L}_{MSE}(\textbf{W}x_i, x_i) ],
\end{align}
However, the original version of \cite{meng2024adversarial} requires the access to all training dataset $\mathcal{D}$ and the control of the training process of the model $f$.
We modify the AdverseMT to only access to the training dataset $\mathcal{D}_{train}$.
Note that the adversarial loss dose not have the special design for multi-target backdoor attacks, we only run the process $c$ times for obtaining $c$ spatial filters for different classes.
So the poisoned subset are $\mathcal{S}_p = \{(\textbf{W}_0(x), 0), (\textbf{W}_1(x), 1), (\textbf{W}_2(x), 2), (\textbf{W}_3(x), 3)\}$.

\subsection{Reinforcement Learning Policy Network Architecture}
Here, we design a concise but effective convolutional neural networks as the our policy network, which is defined as belows: 

\begin{table}[H]
  \setlength\tabcolsep{12pt}
  \caption{The Architecture of Policy Network}
  \label{tab:policynet}
  \centering
  \begin{tabular}{lcccc}
    \toprule
    {Layer} & \multicolumn{1}{c}{In} & \multicolumn{1}{c}{Out} & \multicolumn{1}{c}{Kernel} & Stride \\
    \midrule
    Conv2d & 1 & 32 & (1, 3) & (1, 1) \\
    BatchNorm2d & \\
    ELU & \\
    AvgPool2d &  &  & & (1,2) \\
    \midrule
    Conv2d & 32 & 64 & (1, 3) & (1, 1) \\
    BatchNorm2d & \\
    ELU & \\
    AvgPool2d &  &  & & (1,2) \\
    \midrule
    AdaptiveAvgPool2d &  &  &  & (1, 1) \\
    Flatten & \\
    Linear & 64 & 256 & &  \\
    \bottomrule
  \end{tabular}
\end{table}

\subsection{Target EEG BCIs' Network Architecture}
Three mostly-used EEG BCI models in real-world applications are investigated in our experiments, covering convolutional neural network (CNN) and recurrent neural network (RNN): 1) EEGNet \citep{lawhern2018eegnet}, 2) DeepCNN \citep{schirrmeister2017deep}, 3) LSTM \cite{tsiouris2018long}. Below we detail the architecture of each network. The EEGNet and DeepCNN are almost the same as the original paper (modified a little for cross-subject setting), LSTM comprises an embedding layer, a one-layer LSTM and a linear classifiers.

EEGNet is a compact and concise convolutional network for EEG BCI, having been proven to be effective in a variety of EEG fields with only 3 convolutional layers. DeepCNN is a little bit deeper than EEGNet, which comprises 4 blocks, 5 convolutional layers in total. The LSTM written by us, as demonstrated in Table~\ref{tab:lstm}, is a very shallow network. Our goal is to develop a model-agnostic BA method for EEG modality.

\begin{table}[H]
  \setlength\tabcolsep{12pt}
  \caption{The Architecture of EEGNet}
  \label{tab:eegnet}
  \centering
  \begin{tabular}{lccc}
    \toprule
    {Layer} & \multicolumn{1}{c}{Kernel} & \multicolumn{1}{c}{Input Size} & \multicolumn{1}{c}{Output Size} \\
    \midrule
    16 $\times$ Conv1d & $(C, 1)$ & $C\times T$ & $16\times 1\times T$\\
    BatchNorm & &$16\times 1\times T$& $16\times 1\times T$ \\
    Transpose & & $16\times 1\times T$ & $1\times 16\times T$ \\
    Dropout & 0.25 & $1\times 16\times T$ & $1\times 16\times T$ \\
    \midrule
    4 $\times$ Conv2d & $(2\times 32)$ & $1\times 16\times T$ & $4\times 16\times T$\\
    BatchNorm & &$4\times 16\times T$& $4\times 16\times T$ \\
    Maxpool2D & (2,4) & $4\times 16\times T$ & $4\times 8\times T/4$\\
    Dropout & 0.25 & $4\times 8\times T/4$ & $4\times 8\times T/4$ \\
    \midrule
    4 $\times$ Conv2d & $(8\times 4)$ & $4\times 8\times T/4$ & $4\times 8\times T/4$\\
    BatchNorm & &$4\times 8\times T/4$& $4\times 8\times T/4$ \\
    Maxpool2D & (2,4) & $4\times 8\times T/4$ & $4\times 4\times T/16$\\
    Dropout & 0.25 & $4\times 4\times T/16$ & $4\times 4\times T/16$ \\
    \midrule
    Softmax Regression & & $4\times 4\times T/16$ & Class Number \\
    \bottomrule
  \end{tabular}
\end{table}

\begin{table}[H]
  \setlength\tabcolsep{12pt}
  \caption{The Architecture of DeepCNN}
  \label{tab:deepcnn}
  \centering
  \begin{tabular}{lccc}
    \toprule
    {Layer} & \multicolumn{1}{c}{Kernel} & \multicolumn{1}{c}{Input Size} & \multicolumn{1}{c}{Output Size} \\
    \midrule
    $F_1 \times$ Conv1d & $(1, 32)$ & $C\times T$ & $F_1\times C\times T$\\
    BatchNorm & &$F_1\times 1\times T$& $F_1\times 1\times T$ \\
    $F_1 \times$ Conv1d & $(C, 1)$ & $F_1\times C\times T$ & $F_1\times 1\times T$\\
    BatchNorm & &$F_1\times 1\times T$& $F_1\times 1\times T$ \\
    MaxPooling & (1,2) & $F_1\times 1\times T$ & $F_1\times 1\times T/2$ \\
    Dropout & 0.25 & $F_1\times 1\times T/2$ & $F_1\times 1\times T/2$ \\
    \midrule
    $F_2 \times$ Conv2d & $(1\times 10)$ & $F_1\times 1\times T/2$ & $F_2\times 1\times T/2$\\
    BatchNorm & &$F_2\times 1\times T/2$& $F_2\times 1\times T/2$ \\
    Maxpool2D & (1,2) & $F_2\times 1\times T/2$ & $F_2\times 1\times T/4$\\
    Dropout & 0.25 & $F_2\times 1\times T/4$ & $F_2\times 1\times T/4$ \\
    \midrule
    $F_3 \times$ Conv2d & $(1\times 10)$ & $F_2\times 1\times T/4$ & $F_3\times 1\times T/4$\\
    BatchNorm & &$F_3\times 1\times T/4$& $F_3\times 1\times T/4$ \\
    Maxpool2D & (1,4) & $F_3\times 1\times T/4$ & $F_3\times 1\times T/16$\\
    Dropout & 0.25 & $F_3\times 1\times T/16$ & $F_3\times 1\times T/16$ \\
    \midrule
    $F_4 \times$ Conv2d & $(1\times 4)$ & $F_3\times 1\times T/16$ & $F_4\times 1\times T/16$\\
    BatchNorm & &$F_4\times 1\times T/16$& $F_4\times 1\times T/16$ \\
    Maxpool2D & (1,4) & $F_4\times 1\times T/16$ & $F_4\times 1\times T/64$\\
    Dropout & 0.25 & $F_4\times 1\times T/64$ & $F_4\times 1\times T/64$ \\
    \midrule
    Softmax Regression & & $F_4\times 1\times T/64$ & Class Number \\
    \bottomrule
  \end{tabular}
\end{table}

\begin{table}[H]
  \setlength\tabcolsep{12pt}
  \caption{The Architecture of LSTM, $n$ is the embedding size.}
  \label{tab:lstm}
  \centering
  \begin{tabular}{lcc}
    \toprule
    {Layer} & \multicolumn{1}{c}{Input Size} & \multicolumn{1}{c}{Output Size} \\
    \midrule
    Linear & $C\times T$ & $n\times T$\\
    ReLU & \\
    Linear & $n\times T$ & $n\times T$\\
    \midrule
    LSTM & $n\times T$ & $n\times T$ \\
    \midrule
    Softmax Regression & $n\times T$ & Class Number \\
    \bottomrule
  \end{tabular}
\end{table}

\section{Attack Performance of Professor X}

\subsection{Different Poisoning Rates}

We present the performance of each backdoor attacks' performance under different poisoning rates in Table \ref{tab:diffpoirate}. We can see that our Professor X outperforms other baseline at all poisoning rates, demonstrating the superiority of Professor X. Note that the performance of Professor X on the MI dataset is significantly robust to low poisoning rates, i.e., ASR of 1.000 when $\rho = 0.05$.

{
\begin{table}
  \setlength\tabcolsep{10pt}
  \caption{Clean (/C) and attack (/B) performance with different poisoning rates for Professor X and other baseline methods. The target model is EEGNet for all cases.}
  \label{tab:diffpoirate}
  \centering
  \begin{tabular}{l|lcccccc}
    \toprule
    \multirow{2}*{\makecell[l]{$\rho$}} & Dataset & \multicolumn{2}{c}{Emotion} & \multicolumn{2}{c}{Motor Imagery} & \multicolumn{2}{c}{Epilepsy} \\
    \cmidrule(r){3-4} \cmidrule(r){5-6} \cmidrule(r){7-8}
     ~ & Method &  Clean  & Attack  &  Clean  & Attack &  Clean  & Attack \\
    \midrule
    \multirow{5}*{\rotatebox{90}{0.05}} & PatchMT & 0.390  & 0.333 & 0.281 & 0.791  & 0.449 & 0.365  \\
    ~ & PulseMT & 0.488  & 0.337 & 0.275 & 0.788  & 0.473 & 0.397  \\
    ~ & ComprsMT  & 0.448  & 0.313 & 0.269 & 0.754  & 0.449 & 0.329 \\
    ~ & Professor X & 0.491  & 0.566 & 0.321 & 1.000  & 0.460 & 0.667  \\
    \midrule
    \multirow{5}*{\rotatebox{90}{0.10}} & PatchMT & 0.443  & 0.334 & 0.279 & 0.785  & 0.452 & 0.400  \\
    ~ & PulseMT & 0.445  & 0.394 & 0.281 & 0.796  & 0.486 & 0.591  \\
    ~ & ComprsMT  & 0.509  & 0.323 & 0.270 & 0.778  & 0.446 & 0.337 \\
    ~ & Professor X & 0.541  & 0.718 & 0.320 & 1.000  & 0.452 & 0.734  \\
    \midrule
    \multirow{5}*{\rotatebox{90}{0.15}} & PatchMT & 0.455  & 0.335 & 0.285 & 0.805  & 0.439 & 0.414  \\
    ~ & PulseMT & 0.438  & 0.514 & 0.280 & 0.787  & 0.447 & 0.669  \\
    ~ & ComprsMT  & 0.488  & 0.332 & 0.275 & 0.792  & 0.461 & 0.374 \\
    ~ & Professor X & 0.528  & 0.805 & 0.322 & 1.000  & 0.460 & 0.781  \\
    \midrule
    \multirow{5}*{\rotatebox{90}{0.20}} & PatchMT & 0.481  & 0.334 & 0.277 & 0.816  & 0.461 & 0.451  \\
    ~ & PulseMT & 0.447  & 0.555 & 0.285 & 0.810  & 0.451 & 0.692  \\
    ~ & ComprsMT  & 0.470  & 0.347 & 0.270 & 0.795  & 0.458 & 0.394 \\
    ~ & Professor X & 0.538  & 0.773 & 0.321 & 1.000  & 0.447 & 0.799  \\
    \midrule
    \multirow{5}*{\rotatebox{90}{0.25}} & PatchMT & 0.487  & 0.335 & 0.281 & 0.820  & 0.444 & 0.483  \\
    ~ & PulseMT & 0.466  & 0.701 & 0.275 & 0.815  & 0.431 & 0.684  \\
    ~ & ComprsMT  & 0.493  & 0.335 & 0.269 & 0.800  & 0.462 & 0.427 \\
    ~ & Professor X & 0.551  & 0.836 & 0.325 & 1.000  & 0.447 & 0.834  \\
    \midrule
    \multirow{5}*{\rotatebox{90}{0.30}} & PatchMT & 0.459  & 0.343 & 0.280 & 0.809  & 0.440 & 0.496  \\
    ~ & PulseMT & 0.486  & 0.810 & 0.272 & 0.816  & 0.451 & 0.716  \\
    ~ & ComprsMT  & 0.499  & 0.331 & 0.269 & 0.825  & 0.455 & 0.481 \\
    ~ & Professor X & 0.526  & 0.829 & 0.320 & 1.000  & 0.451 & 0.756  \\
    \midrule
    \multirow{5}*{\rotatebox{90}{0.35}} & PatchMT & 0.437  & 0.341 & 0.285 & 0.805  & 0.448 & 0.510  \\
    ~ & PulseMT & 0.437  & 0.767 & 0.275 & 0.837  & 0.482 & 0.757  \\
    ~ & ComprsMT  & 0.473  & 0.347 & 0.265 & 0.851  & 0.446 & 0.517 \\
    ~ & Professor X & 0.489  & 0.763 & 0.321 & 1.000  & 0.453 & 0.910  \\
    \midrule
    \multirow{5}*{\rotatebox{90}{0.40}} & PatchMT & 0.490  & 0.345 & 0.283 & 0.824  & 0.460 & 0.549  \\
    ~ & PulseMT & 0.454  & 0.771 & 0.270 & 0.825  & 0.439 & 0.443  \\
    ~ & ComprsMT  & 0.464  & 0.361 & 0.269 & 0.865  & 0.437 & 0.450 \\
    ~ & Professor X & 0.528  & 0.849 & 0.323 & 1.000  & 0.477 & 0.944  \\
    \bottomrule
  \end{tabular}
\end{table}
}

\subsection{Hyperparameter Analysis: Frequency and Electrodes Injection Ratio }

We present the performance of each backdoor attacks performance under different rates in Table \ref{tab:hyper_beta} and Table \ref{tab:hyper_gamma}. It can be observed with the increment of $\beta$ and $\gamma$, the attack performance increases. Because the trigger is bigger in clean EEG data.

{
\begin{table}
  \setlength\tabcolsep{10pt}
  \caption{Clean (/C) and attack (/B) performance with frequency injection rate $\beta$, $\gamma = 0.5$}
  \label{tab:hyper_beta}
  \centering
  \begin{tabular}{l|lcccccc}
    \toprule
    \multirow{2}*{\makecell[l]{$\beta$}} & Dataset & \multicolumn{2}{c}{Emotion} & \multicolumn{2}{c}{Motor Imagery} & \multicolumn{2}{c}{Epilepsy} \\
    \cmidrule(r){3-4} \cmidrule(r){5-6} \cmidrule(r){7-8}
     ~ & Method &  Clean  & Attack  &  Clean  & Attack &  Clean  & Attack \\
    \midrule
    \multirow{3}*{\rotatebox{90}{0.05}} & PatchMT & 0.411  & 0.334 & 0.272 & 0.801  & 0.476 & 0.499  \\
    ~ & PulseMT & 0.464  & 0.752 & 0.265 & 0.800  & 0.505 & 0.670  \\
    ~ & Professor X & 0.522  & 0.744 & 0.319 & 0.999  & 0.482 & 0.923  \\
    \midrule
    \multirow{3}*{\rotatebox{90}{0.10}} & PatchMT & 0.431  & 0.363 & 0.283 & 0.824  & 0.482 & 0.540  \\
    ~ & PulseMT & 0.460  & 0.795 & 0.270 & 0.825  & 0.486 & 0.704  \\
    ~ & Professor X & 0.522  & 0.813 & 0.323 & 1.000  & 0.500 & 0.944  \\
    \midrule
    \multirow{3}*{\rotatebox{90}{0.15}} & PatchMT & 0.413  & 0.371 & 0.275 & 0.821  & 0.464 & 0.587  \\
    ~ & PulseMT & 0.449  & 0.701 & 0.271 & 0.821  & 0.477 & 0.632  \\
    ~ & Professor X & 0.532  & 0.848 & 0.322 & 0.998  & 0.477 & 0.947  \\
    \midrule
    \multirow{3}*{\rotatebox{90}{0.20}} & PatchMT & 0.390  & 0.377 & 0.271 & 0.829  & 0.479 & 0.644  \\
    ~ & PulseMT & 0.434  & 0.769 & 0.270 & 0.819  & 0.484 & 0.606  \\
    ~ & Professor X & 0.529  & 0.882 & 0.325 & 0.999  & 0.486 & 0.950  \\
    \midrule
    \multirow{3}*{\rotatebox{90}{0.25}} & PatchMT & 0.406  & 0.385 & 0.267 & 0.835  & 0.491 & 0.673  \\
    ~ & PulseMT & 0.491  & 0.705 & 0.275 & 0.832  & 0.478 & 0.566  \\
    ~ & Professor X & 0.519  & 0.865 & 0.328 & 0.999  & 0.486 & 0.941  \\
    \midrule
    \multirow{3}*{\rotatebox{90}{0.30}} & PatchMT & 0.417  & 0.382 & 0.269 & 0.831  & 0.464 & 0.706  \\
    ~ & PulseMT & 0.425  & 0.708 & 0.273 & 0.844  & 0.488 & 0.592  \\
    ~ & Professor X & 0.521  & 0.862 & 0.330 & 0.999  & 0.495 & 0.940  \\
    \midrule
    \multirow{3}*{\rotatebox{90}{0.35}} & PatchMT & 0.435  & 0.373 & 0.270 & 0.841  & 0.475 & 0.734  \\
    ~ & PulseMT & 0.423  & 0.621 & 0.276 & 0.839  & 0.479 & 0.589  \\
    ~ & Professor X & 0.527  & 0.850 & 0.332 & 0.998  & 0.496 & 0.947  \\
    \midrule
    \multirow{3}*{\rotatebox{90}{0.40}} & PatchMT & 0.438  & 0.378 & 0.271 & 0.843  & 0.469 & 0.751  \\
    ~ & PulseMT & 0.481  & 0.624 & 0.272 & 0.845  & 0.485 & 0.592  \\
    ~ & Professor X & 0.521  & 0.893 & 0.330 & 0.999  & 0.501 & 0.951  \\
    \midrule
    \multirow{3}*{\rotatebox{90}{0.45}} & PatchMT & 0.460  & 0.385 & 0.266 & 0.844  & 0.481 & 0.742  \\
    ~ & PulseMT & 0.429  & 0.633 & 0.277 & 0.856  & 0.499 & 0.601  \\
    ~ & Professor X & 0.519  & 0.877 & 0.325 & 0.999  & 0.492 & 0.962  \\
    \midrule
    \multirow{3}*{\rotatebox{90}{0.50}} & PatchMT & 0.423  & 0.386 & 0.263 & 0.840  & 0.480 & 0.752  \\
    ~ & PulseMT & 0.459  & 0.514 & 0.273 & 0.851  & 0.492 & 0.610  \\
    ~ & Professor X & 0.528  & 0.893 & 0.329 & 1.000  & 0.497 & 0.970  \\
    \bottomrule
  \end{tabular}
\end{table}
}

{
\small
\begin{table}
  \setlength\tabcolsep{10pt}
  \caption{Clean (/C) and attack (/B) performance with electrodes injection rate $\gamma$, $\beta = 0.1$}
  \label{tab:hyper_gamma}
  \centering
  \begin{tabular}{l|lcccccc}
    \toprule
    \multirow{2}*{\makecell[l]{$\gamma$}} & Dataset & \multicolumn{2}{c}{Emotion} & \multicolumn{2}{c}{Motor Imagery} & \multicolumn{2}{c}{Epilepsy} \\
    \cmidrule(r){3-4} \cmidrule(r){5-6} \cmidrule(r){7-8}
     ~ & Method &  Clean  & Attack  &  Clean  & Attack &  Clean  & Attack \\
    \midrule
    \multirow{5}*{\rotatebox{90}{0.10}} & PatchMT & 0.431  & 0.334 & 0.268 & 0.795  & 0.470 & 0.529  \\
    ~ & PulseMT & 0.425  & 0.498 & 0.269 & 0.802  & 0.502 & 0.717  \\
    ~ & ComprsMT  & 0.407  & 0.349 & 0.271 & 0.805  & 0.482 & 0.656 \\
    ~ & Professor X & 0.489  & 0.485 & 0.235 & 0.367  & 0.499 & 0.814  \\
    \midrule
    \multirow{5}*{\rotatebox{90}{0.20}} & PatchMT & 0.473  & 0.335 & 0.271 & 0.805  & 0.464 & 0.599  \\
    ~ & PulseMT & 0.469  & 0.707 & 0.270 & 0.816  & 0.502 & 0.737  \\
    ~ & ComprsMT  & 0.465  & 0.363 & 0.268 & 0.812  & 0.514 & 0.704 \\
    ~ & Professor X & 0.481  & 0.709 & 0.235 & 0.367  & 0.486 & 0.860  \\
    \midrule
    \multirow{5}*{\rotatebox{90}{0.30}} & PatchMT & 0.423  & 0.343 & 0.272 & 0.803  & 0.486 & 0.613  \\
    ~ & PulseMT & 0.488  & 0.767 & 0.273 & 0.814  & 0.506 & 0.749  \\
    ~ & ComprsMT  & 0.451  & 0.398 & 0.271 & 0.811  & 0.494 & 0.700 \\
    ~ & Professor X & 0.500  & 0.743 & 0.235 & 0.367  & 0.490 & 0.883  \\
    \midrule
    \multirow{5}*{\rotatebox{90}{0.40}} & PatchMT & 0.453  & 0.343 & 0.270 & 0.812  & 0.478 & 0.525  \\
    ~ & PulseMT & 0.467  & 0.786 & 0.271 & 0.816  & 0.498 & 0.688  \\
    ~ & ComprsMT  & 0.443  & 0.361 & 0.270 & 0.820  & 0.506 & 0.634 \\
    ~ & Professor X & 0.491  & 0.767 & 0.235 & 0.367  & 0.478 & 0.912  \\
    \midrule
    \multirow{5}*{\rotatebox{90}{0.50}} & PatchMT & 0.431  & 0.363 & 0.270 & 0.813  & 0.472 & 0.552  \\
    ~ & PulseMT & 0.460  & 0.795 & 0.269 & 0.819  & 0.471 & 0.710  \\
    ~ & ComprsMT  & 0.430  & 0.366 & 0.269 & 0.821  & 0.503 & 0.640 \\
    ~ & Professor X & 0.522  & 0.813 & 0.235 & 0.367  & 0.477 & 0.944  \\
    \midrule
    \multirow{5}*{\rotatebox{90}{0.60}} & PatchMT & 0.452  & 0.377 & 0.267 & 0.819  & 0.480 & 0.549  \\
    ~ & PulseMT & 0.460  & 0.808 & 0.269 & 0.823  & 0.490 & 0.672  \\
    ~ & ComprsMT  & 0.459  & 0.368 & 0.271 & 0.826  & 0.499 & 0.534 \\
    ~ & Professor X & 0.488  & 0.828 & 0.235 & 0.367  & 0.495 & 0.950  \\
    \midrule
    \multirow{5}*{\rotatebox{90}{0.70}} & PatchMT & 0.443  & 0.368 & 0.272 & 0.812  & 0.497 & 0.525  \\
    ~ & PulseMT & 0.437  & 0.809 & 0.270 & 0.821  & 0.459 & 0.716  \\
    ~ & ComprsMT  & 0.456  & 0.366 & 0.273 & 0.835  & 0.492 & 0.571 \\
    ~ & Professor X & 0.527  & 0.853 & 0.235 & 0.367  & 0.489 & 0.955  \\
    \midrule
    \multirow{5}*{\rotatebox{90}{0.80}} & PatchMT & 0.461  & 0.383 & 0.268 & 0.821  & 0.479 & 0.573  \\
    ~ & PulseMT & 0.456  & 0.771 & 0.267 & 0.829  & 0.488 & 0.699  \\
    ~ & ComprsMT  & 0.431  & 0.383 & 0.270 & 0.833  & 0.488 & 0.475 \\
    ~ & Professor X & 0.539  & 0.865 & 0.235 & 0.367  & 0.489 & 0.960  \\
    \midrule
    \multirow{5}*{\rotatebox{90}{0.90}} & PatchMT & 0.439  & 0.400 & 0.271 & 0.817  & 0.478 & 0.540  \\
    ~ & PulseMT & 0.461  & 0.811 & 0.269 & 0.823  & 0.494 & 0.694  \\
    ~ & ComprsMT  & 0.459  & 0.389 & 0.274 & 0.836  & 0.490 & 0.309 \\
    ~ & Professor X & 0.520  & 0.824 & 0.235 & 0.367  & 0.489 & 0.970  \\
    \midrule
    \multirow{5}*{\rotatebox{90}{1.00}} & PatchMT & 0.430  & 0.370 & 0.267 & 0.823  & 0.476 & 0.526  \\
    ~ & PulseMT & 0.456  & 0.794 & 0.271 & 0.829  & 0.482 & 0.716  \\
    ~ & ComprsMT  & 0.453  & 0.376 & 0.269 & 0.830  & 0.490 & 0.334 \\
    ~ & Professor X & 0.532  & 0.846 & 0.235 & 0.367  & 0.491 & 0.978  \\
    \bottomrule
  \end{tabular}
\end{table}
}

\subsection{Hyperparameter Analysis in Reinforcement Learning}
\label{apd:rlqt}
We applied the following reward function to acquire the optimal mask strategies for each triggers:
\begin{align}
    Q_t = \mathrm{CA} + \lambda\  \mathrm{ASR} + \mu \ \mathrm{dis}(\mathcal{M}_f^{c_i}) + \nu \min(\mathcal{M}_f^{c_i}),
    \label{Q}
\end{align}
where the first part means the clean accuracy, the second part means the attack success rate, the third part is aiming to scatter the injection positions in various frequency bands, and the fourth part is aiming to inject as high frequencies in EEG signals as possible.
Here, we give a simple example to demonstrate the reward function. For an 10 timepoints long EEG segment $x_i$, $\widetilde{x}_i = \mathcal{F}(x_i)$. If the $\mathcal{M}_f^{c_i} = \{2, 3, 5, 7, 9\}$, because the minimal distance between each pair in $\mathcal{M}_f^{c_i}$ is $|2-3|=1$, thus $\mathrm{dis}(\mathcal{M}_f^{c_i}) = 1$. The $\min(\mathcal{M}_f^{c_i})$ means the lowest position in $\mathcal{M}_f^{c_i}$, thus $\min(\mathcal{M}_f^{c_i}) = 2$.

The analysis of the $\lambda$ are presented in Table \ref{tab:hyper_lambda}. When $\lambda$ increase, the Attack performance increases while the Clean performance declines slightly.

\begin{table}[H]
  \setlength\tabcolsep{5pt}
  \caption{Clean (/C) and attack (/B) performance with ASR's hyperparameter $\lambda$, $\mu = 0.3, \nu=0.005$}
  \label{tab:hyper_lambda}
  \centering
  \begin{tabular}{l|lcccccc}
    \toprule
    \multirow{2}*{\makecell[l]{}} & Dataset & \multicolumn{2}{c}{Emotion} & \multicolumn{2}{c}{Motor Imagery} & \multicolumn{2}{c}{Epilepsy} \\
    \cmidrule(r){3-4} \cmidrule(r){5-6} \cmidrule(r){7-8}
     ~ & Method &  Clean  & Attack  &  Clean  & Attack &  Clean  & Attack \\
    \midrule
    \multirow{1}*{{0.5}} 
    ~ & Professor X & 0.542\scriptsize{$\pm$0.03}  & 0.847\scriptsize{$\pm$0.04} & 0.327\scriptsize{$\pm$0.02} & 1.000\scriptsize{$\pm$0.01}  & 0.500\scriptsize{$\pm$0.04} & 0.922\scriptsize{$\pm$0.04}  \\
    \midrule
    \multirow{1}*{{1.0}} 
    ~ & Professor X & 0.537\scriptsize{$\pm$0.02}  & 0.855\scriptsize{$\pm$0.03} & 0.325\scriptsize{$\pm$0.02} & 1.000\scriptsize{$\pm$0.01}  & 0.482\scriptsize{$\pm$0.03} & 0.935\scriptsize{$\pm$0.05}  \\
    \midrule
    \multirow{1}*{{2}} 
    ~ & Professor X & 0.535\scriptsize{$\pm$0.03}  & 0.857\scriptsize{$\pm$0.02} & 0.323\scriptsize{$\pm$0.02} & 1.000\scriptsize{$\pm$0.01} & 0.477\scriptsize{$\pm$0.04} & 0.944\scriptsize{$\pm$0.02} \\
    \bottomrule
  \end{tabular}
\end{table}

\clearpage
\section{More Visualization Results}
In this section, we plot the reconstructed triggers and masks on three datasets in Section \ref{apd:nc}, more visualizations of backdoor samples in Section \ref{apd:ba_vis}, and the learning curve of our reinforcement learning in Section \ref{apd:rl_vis}.

\subsection{Neural Cleanse: Reconstruction Trigger Patterns}
\label{apd:nc}
\begin{figure}[H]
    \centering
    \includegraphics[width=1.0\textwidth]{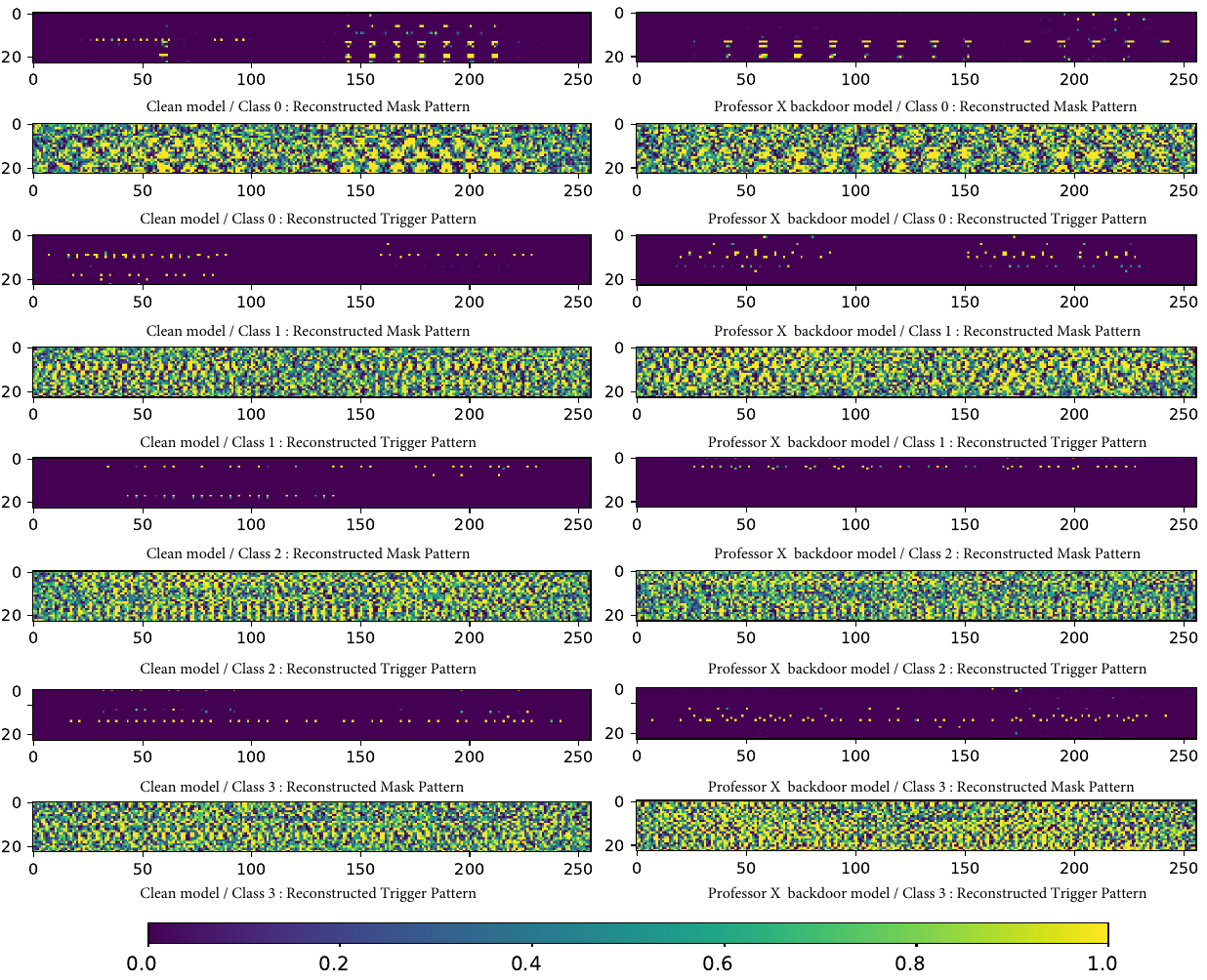}
    \caption{The reconstructed trigger patterns and mask patterns for each possible class in the CHB-MIT dataset. The results in the left column are reconstructed based on the clean model, the results in the right column are reconstructed based on the backdoor model. The EEG segments in the CHB-MIT dataset have 23 electrodes and 256 timepoints.}
    \label{fig:vis_chbmit_nc}
\end{figure}

\begin{figure}[H]
    \centering
    \includegraphics[width=1.0\textwidth]{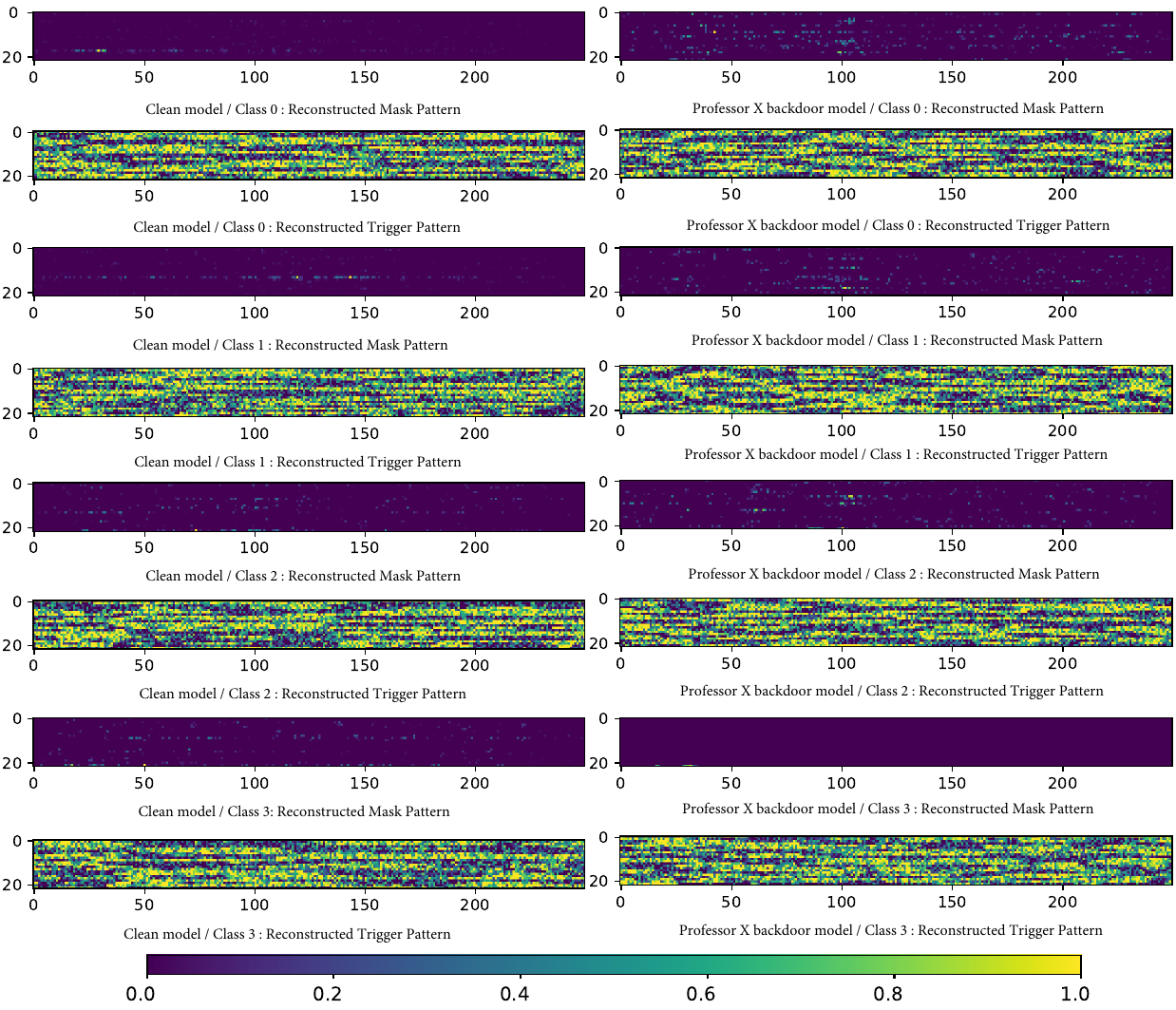}
    \caption{The reconstructed trigger patterns and mask patterns for each possible class in the MI dataset. The results in the left column are reconstructed based on the clean model, the results in the right column are reconstructed based on the backdoor model. The EEG segments in the MI dataset have 22 electrodes and 250 timepoints.}
    \label{fig:vis_mi_nc}
\end{figure}

\begin{figure}[H]
    \centering
    \includegraphics[width=1.0\textwidth]{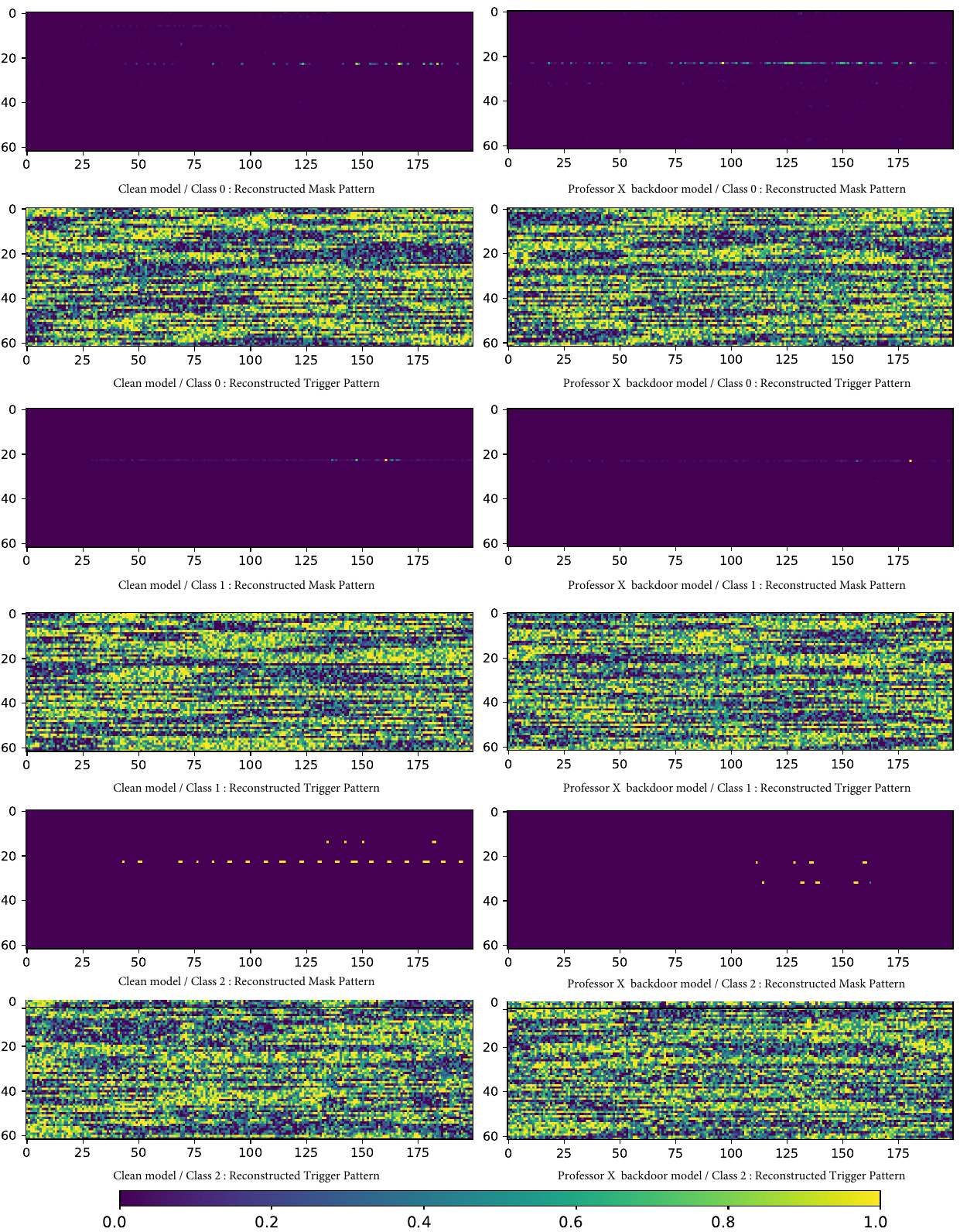}
    \caption{The reconstructed trigger patterns and mask patterns for each possible class in the ER dataset (i.e., SEED dataset). The results in the left column are reconstructed based on the clean model, the results in the right column are reconstructed based on the backdoor model. The EEG segments in the SEED dataset have 62 electrodes and 200 timepoints.}
    \label{fig:vis_er_nc}
\end{figure}

Here, we present more visualization in Figure \ref{fig:vis_chbmit_nc}, Figure \ref{fig:vis_mi_nc}, and Figure \ref{fig:vis_er_nc} of the reconstructed trigger patterns and mask patterns for each possible label on three dataset (\textit{i.e.}, the CHB-MIT dataset, the BCIC-IV-2a dataset and the SEED dataset)  the target model is EEGnet.
It can be observed that the reconstructed trigger patterns and mask patterns of the clean models and Professor X backdoor-injected models are very similar to each other.
Thus, our Professor X backdoor attack can easily bypass the defense of Neural Cleanse.

\newpage
\subsection{Visualization of Backdoor Attack Samples}
\label{apd:ba_vis}
\begin{figure}[H]
    \centering
    \includegraphics[width=1.0\textwidth]{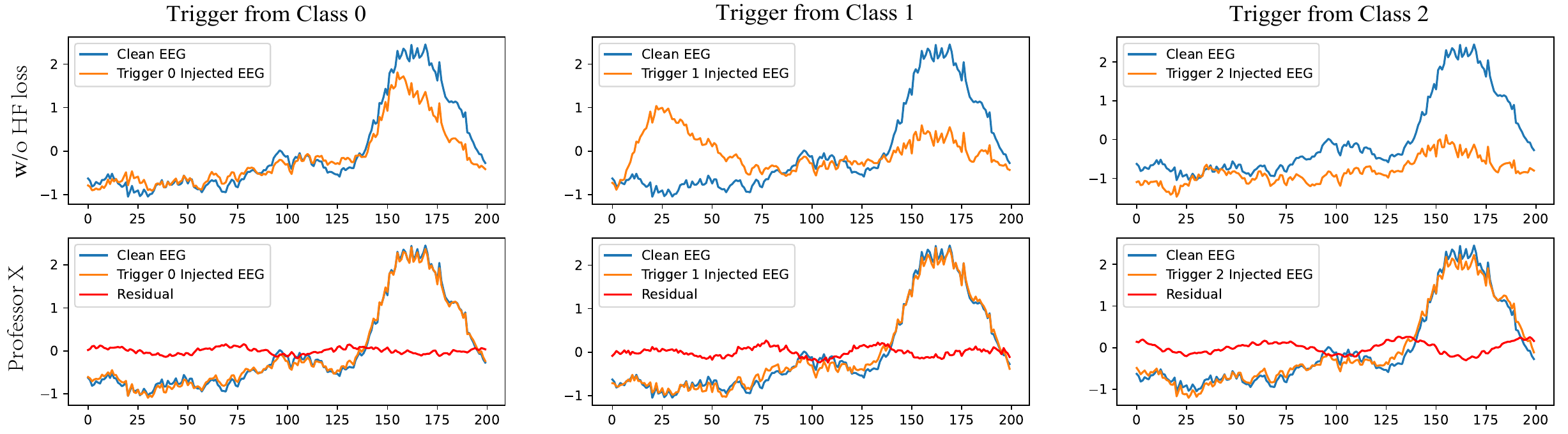}
    \caption{The Clean EEG (Blue), Trigger-injected EEG (Orange) and the Residual (Red) of the ER
dataset.}
    \label{fig:vis_er}
\end{figure}

\begin{figure}[H]
    \centering
    \includegraphics[width=1.0\textwidth]{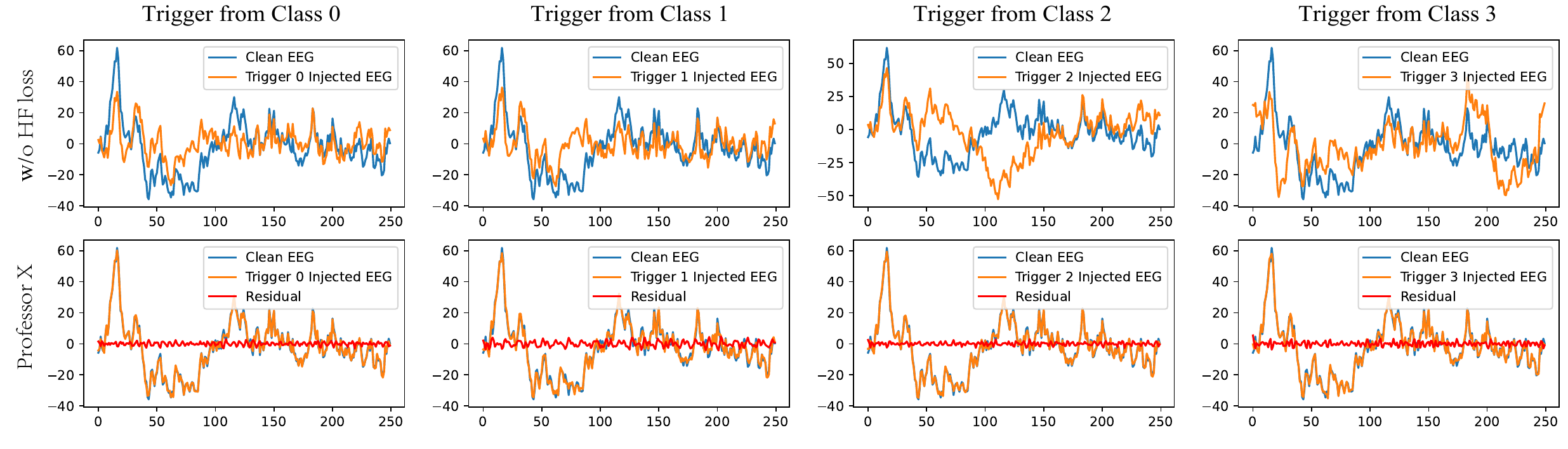}
    \caption{The Clean EEG (Blue), Trigger-injected EEG (Orange) and the Residual (Red) of the MI
dataset.}
    \label{fig:vis_mi}
\end{figure}

We present more visualization of the backdoor attack samples generated by our Professor X on ER
dataset and MI dataset in Fig \ref{fig:vis_er} and \ref{fig:vis_mi}. The x-axis is the timepoints, the y-axis is the normalized
amplitude. Top row: w.o. HF loss; Bottom row: with HF loss. Each column indicates each possible
class.

\newpage
\subsection{Visualization of Learning Curves of Reinforcement Learning}
\label{apd:rl_vis}
\begin{figure}[H]
    \centering
    \includegraphics[width=1.0\textwidth]{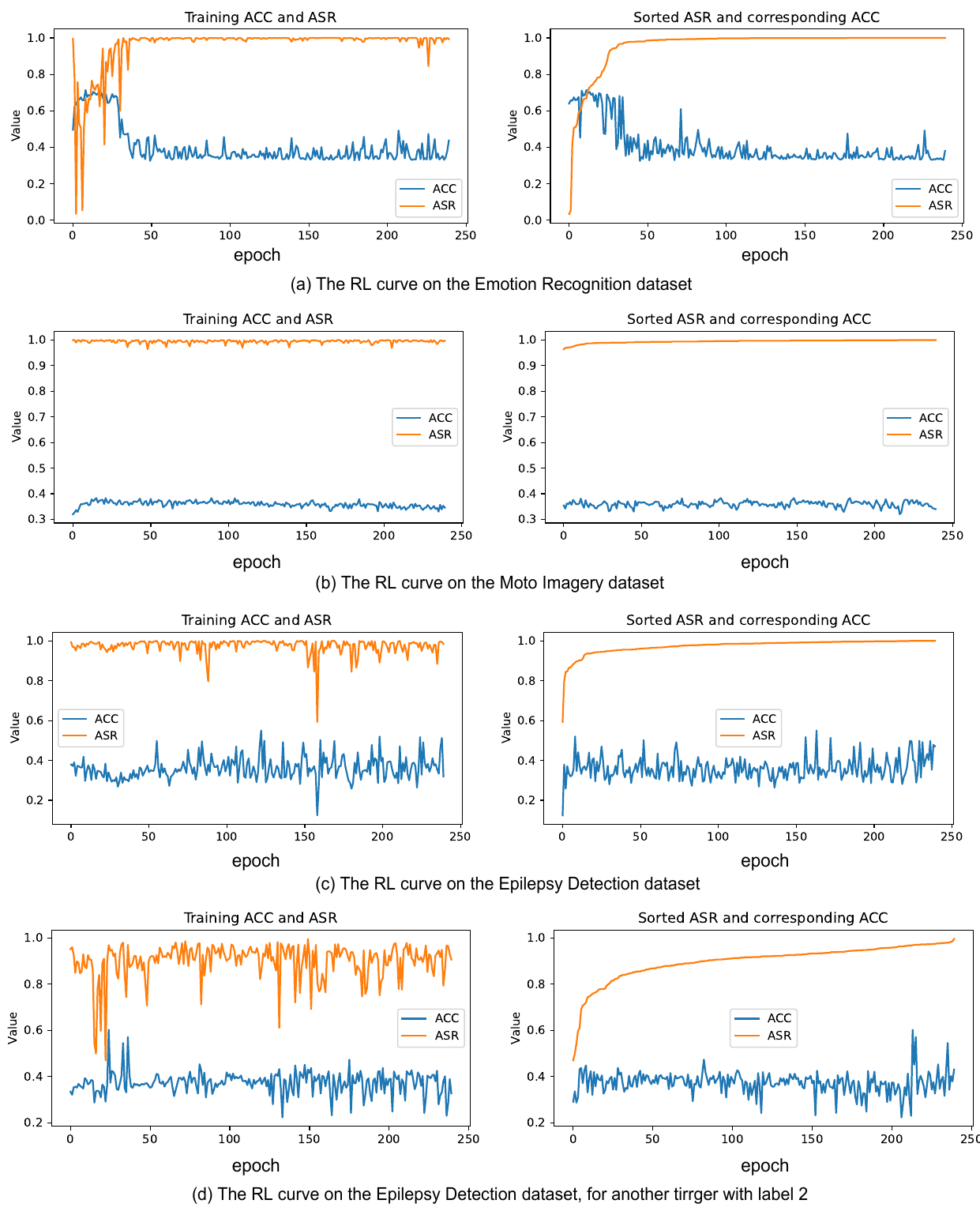}
    \caption{The learning curves of RL on three datasets. The right column is the curve we sort the (Clean Accuracy and Attack Success Rate) (ACC,ASR) according to the ASR. The backdoor models are all EEGNet.}
    \label{fig:rl_vis}
\end{figure}

We present the visualization of the learning curves of the reinforcement learning of three dataset in Fig \ref{fig:rl_vis}. We can see the effectiveness of our reinforcement, which converged within 50 epochs on the ER dataset, that is, only trained 50 backdoor models with different injection strategies.
Our RL is more effective on the MI dataset and ED dataset, which finds a good strategy within less 10 epochs.
Our RL is robust when learning strategies for different triggers as demonstrated in Fig \ref{fig:rl_vis}(c) and (d), where the learning curves are quite similar when RL is performing on different triggers.

\end{document}